\documentclass[preprint, nofootinbib,aps,prd]{revtex4-1}

\usepackage{graphicx}
\usepackage{dcolumn}
\usepackage{bm}
\usepackage{hyperref}
\usepackage{amsmath, amssymb}
\usepackage{slashed}
\usepackage{latexsym}
\usepackage{braket}
\usepackage[export]{adjustbox}
\usepackage{color}
\usepackage{nicematrix}

\begin{document}

\title{Rotating metrics and new multipole moments\\ from scattering amplitudes in arbitrary dimensions}

\author{Claudio Gambino}
\email{claudio.gambino@uniroma1.it}
\author{Paolo Pani}
\email{paolo.pani@uniroma1.it}
\author{Fabio Riccioni}
\email{fabio.riccioni@roma1.infn.it}
\affiliation{Dipartimento di Fisica, Sapienza Universit\`a di Roma \& INFN Roma1, Piazzale Aldo Moro 5, 00185, Roma, Italy}

\begin{abstract}
We compute the vacuum metric generated by a generic rotating object in arbitrary dimensions up to third post-Minkowskian order by computing the classical contribution of scattering amplitudes describing the graviton emission by massive spin-1 particles up to two loops.
The solution depends on the  mass, angular momenta, and on up to two parameters related to generic quadrupole moments.
In $D=4$ spacetime dimensions, we recover the vacuum Hartle-Thorne solution describing a generic spinning object to second order in the angular momentum, of which the Kerr metric is a particular case obtained for a specific mass quadrupole moment dictated by the uniqueness theorem. At the level of the effective action, the case of minimal couplings corresponds to the  Kerr black hole, while any other mass quadrupole moment requires non-minimal couplings.
In $D>4$, the absence of black-hole uniqueness theorems implies that there are multiple spinning black hole solutions with different topology. Using scattering amplitudes, we find a generic solution depending on the mass, angular momenta, the mass quadrupole moment, and a new stress quadrupole moment which does not exist in $D=4$. 
As special cases, we recover the Myers-Perry and the single-angular-momentum black ring solutions,
to third and first post-Minkowksian order, respectively. Interestingly, at variance with the four dimensional case, none of these solutions corresponds to the minimal coupling in the effective action.
This shows that, from the point of view of scattering amplitudes, black holes are the ``simplest'' General Relativity vacuum solutions only in $D=4$.
\end{abstract}

\maketitle

\tableofcontents

\section{Introduction}\label{sec:Introduction}

Despite being non-renormalizable, General Relativity can be treated as a consistent quantum field theory if viewed as the leading-order, low-energy effective field theory arising from a quantum theory of gravity. Specifically, the Einstein-Hilbert action is considered as the initial term in a higher-derivative expansion, where higher-order operators are suppressed at low energy through inverse powers of the Planck mass. Within this view, the gravitational interactions can be computed through the exchange of spin-2 gravitons, giving rise to graviton vertices and matter interactions order by order in a perturbative expansion in $\hbar$, as in ordinary quantum field theories~\cite{Feynman:1963ax,DeWitt:1967yk,DeWitt:1967ub,DeWitt:1967uc,tHooft:1974toh,Donoghue:1993eb,Donoghue:1994dn}.

Remarkably, loop corrections in this expansion give rise not only to Planck-suppressed quantum terms, but also to entirely classical terms that survive in the $\hbar\to0$ limit~\cite{Iwasaki:1971vb,Donoghue:1994dn,Bjerrum-Bohr:2002fji,Donoghue:1996mt,Holstein:2004dn,Bjerrum-Bohr:2018xdl,Kosower:2018adc,Cheung:2018wkq, Guevara:2017csg}. Specifically, the Schwarzschild metric at first order in a post-Minkowskian (i.e., in powers of the gravitational coupling $G$) expansion can be obtained from the scattering amplitude of a massive scalar field emitting gravitons at tree level, while next order post-Minkowskian corrections are obtained by the classical contributions of each $n-$loop amplitude containing graviton vertices\footnote{Recently, the authors of~\cite{Damgaard:2024fqj} managed to resum the post-Minkowskian series and obtain the exact Schwarzschild solution.}~\cite{Bjerrum-Bohr:2002fji}.
This result generalizes to the case of charged and/or spinning geometries by computing the scattering of particles with electric charge and/or spin, reproducing the post-Minkowskian expansion of the Reissner-Nordstrom, Kerr, and Kerr-Newman metrics in $D=4$ spacetime dimensions~\cite{Bjerrum-Bohr:2002fji,Donoghue:2001qc,DOnofrio:2022cvn,Gambino:2022kvb}.

In a more recent development, a systematic method for extracting the classical component of loop amplitudes involving massive scalars interacting with gravitons in any dimension was introduced in~\cite{Bjerrum-Bohr:2018xdl}. This procedure not only demonstrates the agreement of these computations with the earlier work of~\cite{Duff:1973zz} at the second post-Minkowskian order (see also~\cite{KoemansCollado:2018hss,Jakobsen:2020ksu}), but also reveals that the Schwarzschild-Tangherlini metric~\cite{Tangherlini:1963bw} in generic $D$ dimensions at fourth order in the post-Minkowskian expansion emerges from gravitational scattering amplitudes of massive scalars up to three loops~\cite{Mougiakakos:2020laz}.

In this paper we extend this program to the case of spinning geometries both in $D=4$ and $D>4$ dimensions. The motivation for such computation is twofold. 
First, in $D=4$, unlike the case of spherically symmetric spacetimes, the absence of Birkhoff's theorem in axisymmetry implies that the vacuum region outside a spinning object is not necessarily described by the Kerr geometry. Within a post-Minkowskian expansion, the leading-order (linear) angular momentum term is universal, but different spinning objects may have different mass quadrupole moments~\cite{Hartle:1967he, Hartle:1968si,Raposo:2018xkf}. Since the spin-induced quadrupole moment is quadratic in the black-hole spin\footnote{Note that, in this paper, we occasionally refer to ``spin'' both for the spin $s$ of quantum fields and for the angular momentum $J$ of the compact object (e.g., a black hole).}, in order to extend the analysis of~\cite{Bjerrum-Bohr:2002fji} it is necessary to perform computations up to quadratic order in the angular momentum. We shall show that the computation based on the scattering amplitudes provides the post-Minkowskian expansion of the vacuum Hartle-Thorne solution~\cite{Hartle:1967he, Hartle:1968si}, which describes the spacetime of a generic spinning object up to quadratic order in the spin. The Kerr black hole is a particular case of this family, wherein the mass quadrupole moment is fixed by regularity at the horizon and by the black-hole uniqueness theorem~\cite{Israel:1967wq,Carter:1971zc,Hawking:1971vc,Robinson:1975bv,Robinson,Hawking:1973uf,Heusler:1998ua,Chrusciel:2012jk}.
As we shall discuss, at the level of the effective quantum field theory, the Kerr black hole corresponds to the case of minimal couplings, while any other choice of the quadrupole moment requires non-minimal couplings in the action.
The authors of~\cite{Chung:2018kqs, Guevara:2018wpp} proved that in $D=4$ the simplest massive $S$-matrix, defined as the term in the 3-point amplitude which behaves well in the UV, reproduces the Kerr metric in limit in which the spin $s\rightarrow+\infty$. Furthermore, the authors of~\cite{Skvortsov:2023jbn} proved that a spin-s field minimally coupled to gravity also reproduces the dynamics of Kerr black holes, suggesting a correspondence between the simplest massive $S$-matrix and the minimally coupled action in a QFT description for generic spin $s$.
In this sense, one can interpret the minimality of the quantum field theory as the scattering-amplitude counterpart of the celebrated black-hole no-hair theorems in $D=4$ General Relativity~\cite{Hawking:1971vc,Robinson,Hawking:1973uf,Chrusciel:2012jk}.

Our second motivation is that the black-hole uniqueness theorems do not hold in $D>4$~\cite{Emparan:2008eg}, and therefore also in this case it is interesting to compute the metric obtained from the scattering amplitudes and compare it with known solutions.
In particular, spinning black hole solutions in $D>4$ belong to different families and can have different topologies~\cite{Emparan:2008eg}. 
Using scattering amplitudes, we find the generic solution up to third post-Minkowskian order. This solution depends on the mass (\textit{i.e.} mass monopole moment), angular momenta (\textit{i.e.} current dipole moments), mass quadrupole moment and, interestingly, on a new quadrupole moment parameter that we dub \emph{stress quadrupole moment} and is absent in $D=4$.
For specific choices of the parameters, we explicitly check that this general solution reduces to the Myers-Perry black hole~\cite{Myers:1986un} in $D=5$
up to third post-Minkowskian order, and to the black ring~\cite{Emparan:2001wn} with single angular momentum in $D=5$ up to first post-Minkowskian order.
Remarkably, at variance with the four dimensional case, none of these solutions corresponds to the minimal coupling in the effective action. 
This provides strong evidence  that black holes are not the ``simplest'' solutions, from a scattering-amplitude viewpoint, to higher-dimensional General Relativity. A more fundamental explanation for this intriguing result may deserve further investigation.
We explicitly obtain the metric corresponding to the minimal coupling up to third post-Minkowskian order, which is then likely sourced by some matter configuration, similarly to the $D=4$ Hartle-Thorne solution.

In order to obtain the above results we need to compute the metric from the scattering amplitudes of a massive spin-1 field  up to the quadrupole order in a post-Minkowskian expansion. We performed our computations up to two loops (i.e., up to third post-Minkowskian order). The full result is provided in an ancillary Mathematica notebook \cite{AnchillaryFiles}.

Recently Ref.~\cite{Heynen:2023sin} worked out a generalization in $D=5$ of the Thorne formalism~\cite{Thorne:1980ru} for the multipole description in $D=4$ General Relativity.
As we shall discuss in detail, the generalization is based on constructing a suitable coordinate system in which the multipole moments can be read off the asymptotic behavior of the metric components.
Ref.~\cite{Heynen:2023sin} identified the analog of the standard mass and current multipole moments defined by Thorne~\cite{Thorne:1980ru} which are related to the fall-off of the temporal part of the metric.
While we agree with their identification of mass and current moments,  in any $D>4$  we prove the existence of a new multipole moment associated with the asymptotic behavior of the spatial part of the metric. For this reason we call this a \emph{stress} multipole moment, in analogy with the mass and current moments. We expect that in $D>4$ stress multipole moments appear at any order starting from the quadrupolar one, and we define a generalization in arbitrary dimension of the multipole expansion \'a la Thorne, including the new tower of stress moments.
We show that the new stress quadrupole moment is precisely associated to one of the free parameters of our solution and is nonzero already for the Myers-Perry metric.

The rest of the paper is organized as follows. In Sec.~\ref{sec:MetricFromAmplitudes} we review the general approach to compute the metric of a rotating object from scattering amplitudes in arbitrary dimensions.  In Sec.~\ref{sec:MassiveSpin1} we explicitly apply this approach to the scattering of massive spin-1 particles, obtaining the post-Minkowskian metric up to two loops, including the quadrupole moments quadratic in the object angular momentum. This metric is given and discussed in Sec.~\ref{sec:rotmetric} in arbitrary dimensions, with a specific focus on the multipole moments in $D=4$ and $D=5$. In Sec.~\ref{sec:particular} we show that the general solution recovers the cases of the Hartle-Thorne vacuum metric in $D=4$ and the Myers-Perry solution  as well as the single-angular-momentum black ring in $D=5$. Besides, we give an explicit expression for the ``simplest'' metric, {\it i.e.} the one associated with the minimally coupled action, in any dimension. Some details on the multipolar expansion and on these solutions are given in various Appendices. We conclude in Sec.~\ref{sec:conclusions} with a discussion and future prospects.

We work in mostly negative signature with $\eta_{00}=+1$ and in natural units, $\hbar=c=1$, keeping $G\neq 1$. The number of space-time dimensions is $D=d+1$ and the Ricci tensor is defined as $R_{\mu \nu}=R^{\alpha}_{\ \mu \alpha \nu}$. Greek indices $\mu,\nu,...=0, 1, ..., d$ are meant to be contracted either by $g_{\mu\nu}$ or $\eta_{\mu\nu}$ depending on the context, while Latin indices $i, j, ...=1, ..., d$ are contracted with the Euclidean $\delta_{ij}$.

\section{Metric from amplitudes}\label{sec:MetricFromAmplitudes}

In this section we  review the general approach to recover the metric of a rotating object in arbitrary dimension from scattering amplitudes describing the graviton emission by massive spin-$s$ particles~\cite{Mougiakakos:2020laz,DOnofrio:2022cvn}. 
Consider an action of a generic massive spin-$s$ field $\Phi_s$ coupled to gravity\footnote{Notice that we are labelling the massive fields as one does for an object with the same index structure in $d=3$. In higher dimensions more representation are allowed, but this will not be taken into account in our analysis.}
\begin{equation}\label{eq:GenericAction}
    S=\int d^{d+1}x \Bigg(-\frac{2}{\kappa^2}\sqrt{-g}R+
    \mathcal{L}_m(\Phi_s, g_{\mu \nu})\Bigg)\ ,
\end{equation}
where $\kappa^2=32\pi G$. 
Expanding the metric in a post-Minkowskian~(PM) series as
\begin{equation} \label{eq:expansioninh}
    g_{\mu\nu}=\eta_{\mu\nu}+\kappa \, h_{\mu\nu}=\eta_{\mu\nu}+\kappa\sum_{n=1}^{+\infty}h_{\mu\nu}^{(n)}\ ,
\end{equation}
in harmonic gauge the Einstein equations can be rewritten as
\begin{equation}
    \Box h_{\mu\nu}^{(n)}(x)=-\frac{\kappa}{2}\left(T_{\mu\nu}^{(n-1)}(x)-\frac{1}{d-1}\eta_{\mu\nu}T^{(n-1)}(x)\right)\ ,\label{eq:MetricFromEMTxspace}
\end{equation}
where $\Box=\partial_\mu\partial_\nu\eta^{\mu\nu}$ is the flat d'Alambertian operator and $T=\eta^{\mu\nu}T_{\mu\nu}$. In this expression $T_{\mu\nu}^{(0)}(x)$ is the actual stress-energy tensor of the matter source, while $T_{\mu\nu}^{(n)}(x)$ for $n>0$ contain graviton self-interaction terms. 
Then, moving to momentum space, Eq.~\eqref{eq:MetricFromEMTxspace} becomes
\begin{equation}\label{eq:MetricFromEMT}
    h_{\mu\nu}^{(n)}(x)=-\frac{\kappa}{2}\int \frac{d^d \vec q}{(2\pi)^d}\frac{e^{i \vec q\cdot \vec x}}{\vec q\, ^2}\left(T_{\mu\nu}^{(n-1)}(q)-\frac{1}{d-1}\eta_{\mu\nu}T^{(n-1)}(q)\right)\ .
\end{equation}

The idea now is to compute the stress-energy tensor in momentum space by means of 3-point off-shell scattering amplitudes describing the graviton emission from a massive source in the classical limit. The quantization procedure requires the introduction of a gauge-fixing term 
\begin{equation}
    \mathcal{L}_{GF}=\frac{1}{\kappa^2}F^{\lambda}F^{\sigma}\eta_{\lambda \sigma} 
\end{equation}
in the Lagrangian, where $ F^{\lambda}$ is chosen as
\begin{equation}\label{eq:GaugeCondition}
    F^{\lambda}=(1-\alpha)\kappa \, \partial_{\mu}\left(h^{\mu \lambda}-\frac{1}{2}\eta^{\mu \lambda}h\right)+\alpha \, g^{\mu\nu}\Gamma^{\lambda}_{\mu \nu} \ ,
\end{equation}
with $h=\eta^{\mu\nu}h_{\mu\nu}$ and where $\Gamma^{\lambda}_{\mu \nu}$ are the Christoffel symbols. This gauge choice allows us to move continuously from the de Donder gauge ($\alpha=0$) to the harmonic gauge ($\alpha=1$) 
keeping fixed the expression of the graviton propagator~\cite{Jakobsen:2020ksu}
\begin{equation}
    P_{\mu\nu,\rho\sigma} = \frac{1}{2} \left( \eta_{\mu\rho} \eta_{\nu\sigma} + \eta_{\mu\sigma} \eta_{\nu\rho} - \frac{2}{d-1}  \eta_{\mu\nu} \eta_{\rho\sigma} \right) \  .
\end{equation}
 Considering a massive spin-$s$ particle as source of gravitational field, the diagrams that contribute to the classical limit are the ones in which cutting the internal massive lines results in  tree-level diagrams involving only gravitons~\cite{Duff:1973zz,Bjerrum-Bohr:2018xdl}.  

It is therefore possible to compute the classical gravitational conserved current as
\begin{equation}\label{eq:EMTfromAMPpic}
\includegraphics[valign=c,width=0.45\textwidth]{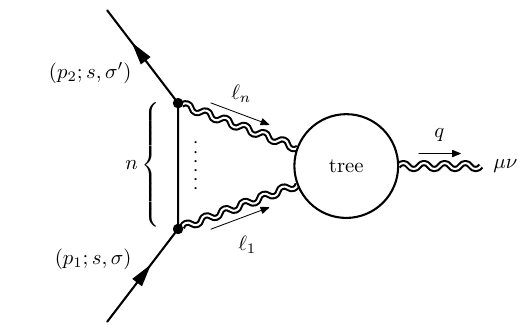}=-\frac{i\, \kappa}{2} (2m)^\epsilon \, T_{\mu\nu}^{(n-1)}(q)\, \delta_{\sigma\sigma'}\ ,
\end{equation}
where $\epsilon=1$ for boson sources and $\epsilon=0$ for fermions\footnote{This is due to the fact that bosons and fermions have a different mass dimension, hence the normalization of the stress-energy tensor is different.}, and where $q=p_1-p_2$ is the transferred momentum. From Eq.~\eqref{eq:EMTfromAMPpic} it is clear that $\sum_{i=1}^n \ell_i = q$ and the order of the PM expansion ($n$) and the one of the loop series ($\ell$) are related by   $n=\ell+1$.  
This can be directly seen from the form of the loop integrals. Indeed, defining  the $l$-loop ``sunset'' master integral
\begin{equation}
    J_{(l)}(\vec q\, ^2)=\int\prod_{i=1}^{l}\frac{d\vec{\ell}_i}{(2\pi)^d}\frac{\vec q\,^ 2}{\left(\prod_{i=1}^{l}\vec{\ell_i}^2\right)\left(\vec q-\vec \ell_1-\cdots-\vec \ell_l\right)^2} \ ,
\end{equation}
as pointed out in~\cite{Mougiakakos:2020laz, DOnofrio:2022cvn}, one can show that $T_{\mu\nu}^{(l)}(q)\propto J_{(l)}(\vec q\, ^2)$, and  
the metric in the long range expansion is given by the Fourier transform
\begin{equation}\label{eq:MasterFT}
    \int\frac{d^d\vec q}{(2\pi)^d}\frac{J_{(l)}(\vec q\, ^2)}{\vec q\, ^2}e^{i\vec q\cdot \vec x}=\left(\frac{\rho(r)}{4\pi}\right)^{l+1}\ ,
\end{equation}
with
\begin{equation}
    \rho(r)=\frac{\Gamma(\frac{d}{2}-1)\pi^{1-d/2}}{r^{d-2}}\ .
\end{equation}
 Notice that we are using a set of Cartesian coordinates in which $x_1^2+\cdots+x_{d}^2=r^2$.

\subsection{Spinning vertices}\label{sec:ClassicalSpin}

Starting from  Eq.\eqref{eq:EMTfromAMPpic}, we have to exploit the classical limit in order to simplify the amplitude calculation. First of all, since we are dealing with spinning particles, the Feynman vertex
\begin{equation}
\includegraphics[valign=c,width=0.30\textwidth]{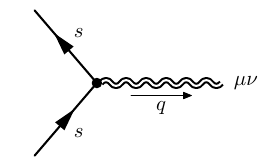}=(\tau_{\Phi^2 h})^{\mu\nu}_{a, b}(q)\ ,
\end{equation}
will have, besides graviton indices $\mu\nu$, spin indices $a$ and $b$, depending on the particular representation of the particles. In the amplitude~\eqref{eq:EMTfromAMPpic}, while the graviton is off-shell, the massive particles are on-shell, meaning that in order to get the stress-energy tensor we have to contract the amplitude with the external polarizations of the particles. From the vertex, at tree level one directly reads the amplitude as
\begin{equation}
    -\frac{i \kappa}{2}(2m)^\epsilon\, T_{\mu\nu}^{(0)}(q)\delta_{\sigma\sigma'}={}^a\bra{p_2;s,\sigma'} (\tau_{\Phi^2 h})_{\mu\nu}^{a, b}\ket{p_1;s, \sigma}^b\ , \label{eq:treelevelamplitude}
\end{equation}
where it is important to notice that since the source particle can be either a fermion or a boson, the polarizations can be described either by a tensor or a spinor, so $a$ and $b$ can be a collection of Lorentz indices or spinor indices. 

Following an approach based on Feynman diagrams~\cite{Bern:2020buy}, which takes into account spin effects in gravity from the classical limit of scattering amplitudes, we now show how to obtain an explicit expression for the tree-level amplitude in Eq.~\eqref{eq:treelevelamplitude}.
To this aim, it is convenient to define
\begin{equation}\label{eq:CapitalP}
    P^\mu = \frac{p_1^\mu+ p_2^\mu }{2}
\end{equation}
so that
\begin{equation}
    p_1^\mu=P^\mu+\frac{1}{2}q^\mu \qquad \text{and} \qquad p_2^\mu=P^\mu-\frac{1}{2}q^\mu\ .
\end{equation}
We are interested in the case in which the source is stationary, meaning that $q^0=0$ and hence the energy is conserved. Then, since $p_1^2=p_2^2=m^2$, just by stationarity one has that $P^\mu=E\, u^\mu=m\, u^{\mu}+O(q)$, where $u^\mu=\delta^\mu_0$ is the velocity of the source, and $E$ is its energy. 
Following~\cite{Kosower:2018adc} to schematically obtain the classical limit out of scattering amplitude calculations, for a process involving a particle with spin $s$ and  transferred momentum $q$, the algorithm is to write the amplitude in terms of the spin tensor ${ S}_{\mu\nu}$, and then make the replacement
\begin{equation}
    q\rightarrow \hbar\,  q \qquad \text{and} \qquad { S}\rightarrow\frac{1}{\hbar}{ S} \ , \label{eq:KMOC}
\end{equation}
keeping only the terms $O(\hbar^0)$, \textit{i.e.} those that survive in the limit $\hbar\rightarrow 0$. Therefore, terms $O(q^n)$ with $n>0$ are quantum, while terms $O(({ S} q)^n$) are classical. Crucially, the presence of the spin compensates the transferred momentum and gives rise to classical terms with higher powers in $q$. 
Notice that  since the two external momenta are equal up to the transferred momentum, using the results in~\cite{Bern:2020buy, Chung:2018kqs} it turns out that in the stationary limit  we can expand the polarization state as 
\begin{equation}\label{eq:ClassicalLimitPolarization}
\ket{p_2}=\ket{p_1}+O(\hbar)\ ,
\end{equation}
and then define the dressed vertex 
\begin{equation}\label{eq:DressedVertexGeneric}
   \bra{p_2; s, \sigma'}(\tau_{\Phi^2 h})^{\mu\nu}\ket{p_1; s, \sigma}=\bra{p_1; s, \sigma'}(\tau_{\Phi^2 h})^{\mu\nu}\ket{p_1; s, \sigma}+O(\hbar)=\hat{\tau}_{\Phi^2 h}^{\mu\nu}(q,S)\delta_{\sigma\sigma'}+O(\hbar)\ ,
\end{equation}
where  the spin-index structure is left understood. 

Now, in order to extract such classical spin pieces, let us consider a field in a particular representation of the Lorentz group, satisfying the algebra 
\begin{equation}
    [M^{\mu \nu}, M^{\rho \sigma}]=-i\Big(\eta^{\mu\rho}M^{\nu\sigma}-\eta^{\nu \rho}M^{\mu \sigma}+\eta^{\nu \sigma}M^{\mu \rho}-\eta^{\mu\sigma}M^{\nu\rho}\Big)\ ,
\end{equation}
with $M$ the generators of the group. Considering then  ``in'' and ``out'' states of spin $s$ and polarizations $\sigma$ and $\sigma'$, generalizing the argument in~\cite{Bern:2020buy} to arbitrary dimension, in the stationary limit we have
\begin{equation}\label{eq:ClassicalLimitOfGenerators}
\begin{aligned}
    \braket{p_2;s, \sigma'|p_1;s, \sigma}&=\braket{p_1;s, \sigma'|p_1;s, \sigma}+O(\hbar)=C(s)\delta_{\sigma\sigma'}\ ,\\
    \braket{p_2;s, \sigma'|M^{\mu\nu}|p_1;s, \sigma}&=S^{\mu\nu}C(s)\delta_{\sigma\sigma'}+O(\hbar^0)\ ,\\
    \braket{p_2;s, \sigma'|\frac{1}{2}\{M^{\mu\nu}, M^{\rho\sigma}\}|p_1;s, \sigma}&=S^{\mu\nu}S^{\rho\sigma}C(s)\delta_{\sigma\sigma'}+O(\hbar^{-1})\ ,
\end{aligned}
\end{equation}
where $\{\cdot,\cdot\}$ stands for the anti-commutator, the external state can be either a fermion or a boson, $C(s)$ is the normalization of the states which depends on the particular representation of the Lorentz group, and we now interpret $S^{\mu\nu}$ as the spin tensor of the classical source of the gravitational field.
It is important to observe that the stationary limit defined above corresponds to the source being in the rest frame, hence in this frame the temporal part of the spin tensor  vanishes, \textit{i.e.} $S^{0i}=0$. 

Finally, we observe that for a  massive particle with spin $s$ the dressed vertex contains powers  up to $2s$ of the spin tensor.  
As we will see, this implies that the resulting metric describes an object with only the first $2s$ multipoles turned on.

\subsection{Loop amplitudes for any spin}\label{sec:LoopSpin}

We now want to discuss how, starting from the analysis above, one can construct loop amplitudes in such a way that the spin structure arising from the massive spin-$s$ particles  is entirely given as a tensor  dependence of the dressed vertex. More specifically, in the classical limit the massive line of Eq.~\eqref{eq:EMTfromAMPpic} factorizes in the product of $n$ dressed vertices in such a way that the loop calculation is formally the one of a scalar process. We will first discuss explicitly the 1-loop case, and then naturally extend it to all loops. 

The contribution to the amplitude arising from the massive line in  Eq.~\eqref{eq:EMTfromAMPpic} for $n=2$ is
\begin{equation}\label{eq:MatterLine1Loop}
    \bra{p_2; s, \sigma'}(\tau_{\Phi^2 h})^{\mu\nu}(\ell) \frac{i\, \mathcal{P}(p_1-\ell)}{(p_1-\ell)^2-m^2+i\varepsilon}(\tau_{\Phi^2 h})^{\rho\lambda}(q-\ell)\ket{p_1; s,\sigma}\ ,
\end{equation}
where $\mathcal{P}$ is the numerator of the matter propagator which depends on the spin of the massive field. For a massive spin-$s$ particle, we can always rewrite the operator $\mathcal{P}$ in terms of the sum of the polarization states as 
\begin{equation}
    \mathcal{P}(p_1-\ell)=(2m)^{1-\epsilon}\sum_{\sigma''}\ket{p_1-\ell;s, \sigma''}\bra{p_1-\ell;s , \sigma''}\ .
\end{equation}
Plugging this back in Eq.~\eqref{eq:MatterLine1Loop} we end up with
\begin{equation}
   \frac{i\, (2m)^{1-\epsilon}}{(p_1-\ell)^2-m^2+i\varepsilon}  \sum_{\sigma''}\bra{p_2;s,\sigma'} (\tau_{\Phi^2 h})^{\mu\nu}(\ell)\ket{p_1-\ell;s, \sigma''}\bra{p_1-\ell;s , \sigma''}(\tau_{\Phi^2 h})^{\rho\lambda}(q-\ell)\ket{p_1; s, \sigma}\ .
\end{equation}
Finally, using the classical limit as in Eq.~\eqref{eq:ClassicalLimitPolarization} we get
\begin{equation}
   \frac{i\, (2m)^{1-\epsilon}}{(p_1-\ell)^2-m^2+i\varepsilon}  \sum_{\sigma''}\bra{p_1;s,\sigma'} (\tau_{\Phi^2 h})^{\mu\nu}(\ell)\ket{p_1;s, \sigma''}\bra{p_1;s , \sigma''}(\tau_{\Phi^2 h})^{\rho\lambda}(q-\ell)\ket{p_1; s, \sigma}+O(\hbar)\ ,
\end{equation}
from which we derive the expression written in terms of the dressed vertices as
\begin{equation}
   \frac{i\, (2m)^{1-\epsilon}}{(p_1-\ell)^2-m^2+i\varepsilon} \hat{\tau}_{\Phi^2 h}^{\mu\nu}(\ell, S)\hat{\tau}_{\Phi^2 h}^{\rho\lambda}(q-\ell,S)\delta_{\sigma \sigma'}\ .
\end{equation}

To summarize, after this dressing procedure the classical limit of a spinning 3-point 1-loop amplitude in the stationary regime completely reduces to an amplitude in which the spin contribution arises  solely from the tensor structure of  the dressed vertex, up to a normalization factor depending on whether one considers fermions or bosons. This can be naturally generalized to any loop, which means that we can use all the machinery of~\cite{Mougiakakos:2020laz, DOnofrio:2022cvn} in order to write down the $l$-loop amplitude related to the stress-energy tensor, which is 
\begin{equation}\label{eq:EMTfromAMPfinal}
    -\frac{i\, \kappa}{2}(2m)T_{\mu\nu}^{(l)}(q)=\frac{(-i)^{l+1}}{(l+1)!}\int\prod_{i=1}^{l}\frac{d^d \vec{\ell_i}}{(2\pi)^d}\frac{\prod_{i=1}^{l+1}\hat{\tau}_{\Phi^2h}^{\mu_i\nu_i}(\ell_i, S)\prod_{i=1}^{l+1}P_{\mu_i\nu_i, \alpha_i\beta_i}}{\prod_{i=1}^{l+1}\vec{\ell_i}^2}\mathcal{M}^{\alpha_1\beta_1, ...,\mu\nu}\ ,
\end{equation}
where $\mathcal{M}^{\alpha_1\beta_1, ..., \alpha_n\beta_n,\mu\nu}$ contains the sum over all the tree-level graviton diagrams as shown in Eq.~\eqref{eq:EMTfromAMPpic}, and in our signature $\ell_i^2=-\vec \ell_i^2$, since it is possible to show that $\ell_i^0=0$ in the classical and stationary limit.

From  Eq.~\eqref{eq:EMTfromAMPfinal} we can systematically compute the stress-energy tensor, and inserting it back into Eq.~\eqref{eq:MetricFromEMT} we can read off the metric
at any PM order $n$. Since the dressed vertex contains powers up to $2s$ in the spin tensor and an $l$-loop amplitude contains $n$ dressed vertices, as can be seen from Eq.~\eqref{eq:EMTfromAMPfinal},   this implies that the highest power of $S$ in the stress-energy tensor is 
 $2ns$.
Hence, the resulting metric at PM order $n$ is expanded as
\begin{equation}\label{eq:MultipoleMetricExpansion}
    h_{\mu\nu}^{(n)} = \sum_{j=0}^{2ns }h_{\mu\nu}^{(n,j)} \ ,
\end{equation}
where $j$ is the order of the expansion in powers of $S$\footnote{For instance, $h_{\mu\nu}^{(1, 1)}$ will be 1PM and linear in the spin, $h_{\mu\nu}^{(1, 2)}$ will be 1PM and quadratic in the spin, and so on.}. 
This is  consistent with the fact that all multipoles starting from $2s+1$ order vanish. 

    What remains to be done in this procedure is to determine  the explicit  expression of the dressed vertex $\hat{\tau}_{\mu\nu}$ for a given spin-$s$ field. 
In the next subsection we will show as an example how the $O(S)$ terms  are derived considering  spin-1/2 spinors, while in the next section we will construct the dressed vertex for a spin-1 field.  

\subsection{Spin-1/2 case}\label{sec:ExampleSpin1/2}

Here we review the work of~\cite{Bjerrum-Bohr:2002fji}  and we extend it to arbitrary dimensions by employing the formalism discussed in the last section. Let us consider a massive spin-1/2 particle as source of gravitational field. Its action reads
\begin{equation}
    S=\int d^4 x\, e\, \overline{\psi}\left(i\, e^\mu_{\ a}\gamma^aD_\mu-m\right)\psi\ ,\label{eq:Diracaction}
\end{equation}
where $g_{\mu\nu}=e^a_{\ \mu}e_{a\nu}$ and
\begin{equation}
D_\mu\psi=\partial_{\mu} \psi+\frac{1}{4}\omega_{\mu ab}\gamma^{ab}\psi\ ,
\end{equation}
with $M^{ab}=\frac{i}{2}\gamma^{ab}$ generators of the Lorentz group in the spin-1/2 representation, $\omega_{\mu ab}$ the spin connection, and $a$ and $b$ flat indices. 

The action in~\eqref{eq:Diracaction} gives rise to the tri-linear interaction vertex
\begin{equation}
   \tau^{\mu\nu, \alpha\beta}_{\psi^2h}(q)=-\frac{i\, k}{2}\Bigg(\frac{1}{4}\gamma^\mu (p_1+p_2)^\nu+\frac{1}{4}\gamma^\nu(p_1+p_2)^\mu-\eta^{\mu\nu}\left(\frac{i}{2} (\slashed{p_1}+\slashed{p_2})-m\right)\Bigg)^{\alpha \beta}\ ,
\end{equation}
  which can be rewritten using the definition in~\eqref{eq:CapitalP} as
\begin{equation}
   \tau^{\mu\nu, \alpha\beta}_{\psi^2h}(q)=-\frac{i\, k}{2}\Bigg(\frac{1}{2}\gamma^\mu P^\nu+\frac{1}{2}\gamma^\nu P^\mu-\eta^{\mu\nu}\left(i \slashed{P}-m\right)\Bigg)^{\alpha \beta}\ ,
\end{equation}
where $\alpha$ and $\beta$ are spinor indices. We now want to  compute the dressed vertex. Using the Dirac equations in momentum space 
\begin{equation}
    \overline{u}(p_2, \sigma')\left(\slashed{P}-m\right)u(p_1, \sigma)=0\ ,
\end{equation}
where $u$ and $\overline{u}$ are the spinor polarizations, we obtain 
\begin{equation}
    \hat{\tau}_{\psi^2h}^{\mu\nu}(q, S)\, \delta_{\sigma\sigma'}=\overline{u}(p_2, \sigma')\tau_{\psi^2h}^{\mu\nu}(q)u(p_1, \sigma)=-\frac{i\, \kappa}{2}\overline{u}(p_2, \sigma')\Big(\frac{1}{2}\gamma^\mu P^\nu+\frac{1}{2}\gamma^\nu P^\mu\Big)u(p_1, \sigma)\ ,
\end{equation}
 where we keep understood the spinor indices. Now, since we are dealing with on-shell massive states, we can simplify the vertex by means of the Gordon identity, which reads
\begin{equation}
    \overline{u}(p_2, \sigma')\gamma^{\mu}u(p_1, \sigma)=\overline{u}(p_2, \sigma')\Big(\frac{1}{m}P^\mu-\frac{i}{m}M^{\mu\nu}q_\nu\Big)u(p_1, \sigma)\ ,
\end{equation}
and finally we get
\begin{equation}
    \hat{\tau}_{\psi^2h}^{\mu\nu}(q, S)\, \delta_{\sigma\sigma'}=-\frac{i\, \kappa}{2}\overline{u}(p_2, \sigma')\left(\frac{1}{m}P^\mu P^\nu-\frac{i}{2 m}q_\lambda(M^{\mu\lambda}P^\nu+M^{\nu\lambda}P^\mu)\right)u(p_1, \sigma)\ .
\end{equation}
Then, using the relations established in~\eqref{eq:ClassicalLimitOfGenerators} with the normalization $C(1/2)=1$, we rewrite the dressed vertex in terms of the classical spin tensor as
\begin{equation}\label{eq:spin1/2ClassicVertex}
     \hat{\tau}_{\psi^2h}^{\mu\nu}(q, S)=-\frac{i\, \kappa}{2}\left(\frac{1}{m}P^\mu P^\nu-\frac{i}{2 m}q_\lambda(S^{\mu\lambda}P^\nu+S^{\nu\lambda}P^\mu)\right)\ .
\end{equation}
Finally, in the stationary limit in which $P^\mu=m\, \delta^\mu_0+O(\hbar)$, the resulting  vertex reads
 \begin{equation}
     \hat{\tau}_{\psi^2h}^{\mu\nu}(q, S)=-\frac{i\, \kappa}{2}\left(m\, \delta^\mu_0 \delta^\nu_0-\frac{i}{2}q_\lambda(S^{\mu\lambda}\delta^\nu_0+S^{\nu\lambda}\delta^\mu_0)\right)\ .
\end{equation}
Notice that the dressed vertex defined above holds in arbitrary dimensions since we have never used any explicit representation of the gamma matrices. 

From the vertex,  it is now straightforward to obtain the metric. At tree-level, the stress-energy tensor  reads
\begin{equation}
    T^{\mu\nu}_{(0)}(q)=\frac{2\, i}{\kappa}\hat{\tau}_{\psi^2h}^{\mu\nu}(q, S)=m\, \delta^\mu_0 \delta^\nu_0-\frac{i}{2}q_\lambda(S^{\mu\lambda}\delta^\nu_0+S^{\nu\lambda}\delta^\mu_0)\ ,
\end{equation}
and using Eqs.~\eqref{eq:MetricFromEMT} and~\eqref{eq:MasterFT} we get for the scalar part of the metric
\begin{equation}\label{eq:MonopoleMetric}
    \begin{aligned}
        h_{00}^{(1,0)}(r)&=-\frac{4(d-2)}{d-1} Gm\, \rho(r)\ ,\\
        h_{0i}^{(1,0)}(r)&=0\ ,\\
        h_{ij}^{(1,0)}(r)&=-\frac{4\delta_{ij}}{d-1}Gm\, \rho(r)\ ,
    \end{aligned}
\end{equation}
which is exactly the Schwarzschild-Tangherlini metric at 1PM, and for the dipole part we obtain
\begin{equation}\label{eq:DipoleMetric}
    \begin{aligned}
        h_{00}^{(1,1)}(r)&=0\ ,\\
        h_{0i}^{(1,1)}(r)&=-\frac{2(d-2)x^k S^{i}{}_{k}}{r^2}G\,  \rho(r)\ ,\\
        h_{ij}^{(1,1)}(r)&=0\ .\\
    \end{aligned}
\end{equation}
Notice that while Greek indices are meant to be contracted by the Minkowski metric, from now on repeated Latin indices will indicate an Euclidean contraction. This means in particular that there is no difference between upper and lower Latin indices.

Finally, we know from General Relativity that at dipole order, the expansion in the far field limit is unique, and therefore we expect the metric 
to be independent of any arbitrary coefficients other than the spacetime mass and the spin tensor. From an amplitude perspective, this means that there are no additional non-minimal couplings that can modify the classical vertex at dipole order. In the next section we will consider a massive spin-1 field, and we will construct the non-minimal effective action giving rise to the most general vertex at quadrupole order. This analysis, that can be extended to fields of arbitrary spin,  will consistently show that there are no non-minimal terms linear in $S$ that can modify the vertex.

\section{Spin-1 scattering amplitudes in arbitrary dimension}\label{sec:MassiveSpin1}

In this section we will compute the metric arising from the graviton emission by a massive spin-1 field. This produces terms up to quadrupole order in the multipole expansion of~\eqref{eq:MultipoleMetricExpansion}. We know from General Relativity that these are the first terms in the expansion that discriminate between different solutions of the Einstein equation. Therefore, from our amplitude perspective, we expect that the dressed vertex will not be uniquely defined, corresponding to the presence of non-minimal couplings in the action. We will first compute the dressed vertex arising from the minimally coupled action, and then we will include the contribution of non-minimal couplings, resulting in the most general stress-energy tensor up to quadrupole order. Finally, we will compute the resulting metric up to 2 loops.

\subsection{Minimal vertex}

Let us consider a massive spin-1 field minimally coupled to gravity, also known as Proca field
\begin{equation}\label{eq:ProcaAction}
    S_{\text{min}}=\int d^{d+1}x\sqrt{-g}\left(-\frac{1}{4}F_{\mu\nu}F^{\mu\nu}+\frac{1}{2}m^2V_{\mu}V^{\mu}\right)\ ,
\end{equation}
where $F_{\mu\nu}=\partial_{\mu}V_{\nu}-\partial_{\nu}V_\mu$ is the usual anti-symmetric strength tensor.
Our aim now is to compute the minimal dressed vertex associated to the action~\eqref{eq:ProcaAction}. In order to do this, consider the tri-linear vertex associated to this action~\cite{Bjerrum-Bohr:2014lea}
\begin{equation}\label{eq:spin1VertexOriginal}
    \begin{aligned}
   \varepsilon_{\beta}(p_2) {\tau}_{V^2h, \text{min}}^{\mu\nu, \beta\alpha}\varepsilon_{\alpha}(p_1)&=-\frac{i\, \kappa}{2}\Bigg(\varepsilon(p_1)\cdot p_2\Big(p_1^\mu\varepsilon^\nu(p_2)+p_1^\nu\varepsilon^\mu(p_2)\Big)+\varepsilon(p_2)\cdot p_1\Big(p_2^\mu\varepsilon^\nu(p_1)+p_2^\nu\varepsilon^\mu(p_1)\Big)\\
    &-\varepsilon(p_1)\cdot\varepsilon(p_2)\Big(p_1^\mu p_2^\nu+p_2^\mu p_1^\nu\Big)-\Big(p_1\cdot p_2-m^2\Big)\Big(\varepsilon(p_1)^\mu\varepsilon(p_2)^\nu+\varepsilon(p_2)^\mu\varepsilon(p_1)^\nu\Big)\\
    &+\eta^{\mu\nu}\Big(\big(p_1\cdot p_2-m^2\big)\varepsilon(p_1)\cdot\varepsilon(p_2)-p_1\cdot\varepsilon(p_2)p_2\cdot\varepsilon(p_1)\Big)\Bigg)\ ,
    \end{aligned}
\end{equation}
where $\varepsilon_{\alpha}(p)$ are the real polarization vectors satisfying the transversality condition $\varepsilon(p)\cdot p=0$, due to the on-shellness of the massive states. 

We now have to manipulate the above expression in order to write it in terms of the Lorentz generators, which in this particular representation take the form
\begin{equation}
    M^{\mu\nu, \rho \sigma}=i \big( \eta^{\mu\rho}\eta^{\nu\sigma}-\eta^{\mu \sigma}\eta^{\nu \rho}\big)\ ,
\end{equation}
which holds in arbitrary dimensions. 
The strategy  is to strip off the polarization vectors from Eq.~\eqref{eq:spin1VertexOriginal} and massage the Lorentz structure in order to rewrite everything in terms of $M^{\mu\nu, \rho \sigma}, q^\mu$ and $P^\mu$. 
Up to terms proportional to  $p_1^\alpha$ and $p_2^{\beta}$, that vanish when contracted with the polarization vectors, 
from Eq.~\eqref{eq:spin1VertexOriginal} one thus obtains
\begin{equation}
    \begin{aligned}    
       \varepsilon_{\beta}(p_2) \tau_{V^2h,\text{min}}^{\mu\nu, \beta\alpha}&\varepsilon_{\alpha}(p_1)=\frac{i\, \kappa}{2} \varepsilon_{\beta}(p_2) \Bigg(\eta^{\alpha\beta}\left(2P^\mu P^\nu+\frac{1}{2}\eta^{\mu\nu}q^2-\frac{1}{2}q^\mu q^\nu\right)\\
    &-iq_\lambda\left(P^\mu M^{\nu\lambda, \beta\alpha}+P^\nu M^{\mu\lambda, \beta\alpha}\right)-q_{\rho}q_{\sigma}\frac{1}{2}\{M^{\mu\rho}, M^{\nu\sigma}\}^{\beta\alpha}\Bigg) \varepsilon_{\alpha}(p_1)\ ,
    \end{aligned}
\end{equation}
from which, employing the relations in~\eqref{eq:ClassicalLimitOfGenerators} and the definition in Eq.~\eqref{eq:DressedVertexGeneric}, the minimal dressed vertex in the stationary limit reads
\begin{equation}\label{eq:MinimalVertex}
    \hat{\tau}_{V^2h, \text{min}}^{\mu\nu}(q)=-\frac{i\, \kappa}{2}\Big(2 m^2 \delta^\mu_0 \delta^\nu_0- i\, m\, q_\lambda \left(S^{\mu\lambda}\delta^{\nu}_0+S^{\nu\lambda}\delta_0^{\mu}\right)-q_\lambda q_\sigma S^{\mu\lambda}S^{\nu\sigma}\Big)\ ,
\end{equation}
where we have normalized the polarization vectors as $C(1)=-1$.
We can see that the scalar and dipole terms in Eq.~\eqref{eq:MinimalVertex}, up to the normalization factor, exactly coincide with Eq.~\eqref{eq:spin1/2ClassicVertex}, as expected from the fact that such terms are uniquely fixed in the metric. 
The remaining terms are $O(S^2)$ and constitute the quadrupole contribution for the minimally coupled spin-1 field. 

\subsection{Non-minimal vertex}

Once computed the dressed vertex associated to the Proca action, it is natural to consider what happens for a non-minimally coupled theory. The effective action will be constructed with several non-minimal terms with unconstrained couplings, that  we expect to enter in the expression of the general dressed vertex as unfixed numerical parameters. In general such effective action is more involved than the minimally-coupled one, and therefore the tri-linear vertex associated to it will be more complicated with respect to~\eqref{eq:spin1VertexOriginal}. This means that carrying out the procedure outlined in the last subsection to get the dressed vertex is not viable. 

Following the idea of~\cite{Bern:2020buy}, we can  define a covariant generalization of the spin tensor in order to build non-minimal operators with an explicit spin dependency, instead of stripping it out as we did in the last subsection. To this end, we define $\mathbb{S}^{\mu\nu}$, a well defined covariant anti-symmetric tensor, such that
\begin{equation}\label{eq:DefinitionSpinTensor}
    \mathbb{S}^{\mu\nu}_{a,b}=S^{\mu\nu}\delta_{ab}+O(\kappa)\ ,
\end{equation}
where $a$ and $b$ are the spin indices depending on the representation of the field over which such operator acts. Before constructing the non-minimal effective action, we  notice that since we are only interested in the long range pieces of the metric, every local term can be neglected. In particular, this means that at tree-level 
we can neglect terms that contain the squared modulus of the transferred momentum, \textit{i.e.} $T_{\mu\nu}^{(0)}(q)=O(|\vec q\,| ^2)$, since they lead to local terms in the metric. This implies that also the dressed vertex is defined up to terms $\hat{\tau}_{\mu\nu}(q)=O(|\vec q\,| ^2)$. 

Now we can construct the non-minimal effective action as
\begin{equation}\label{eq:NonMinimalAction}
    \begin{aligned}
    S_{\text{non-min}}&=\int d^Dx \sqrt{-g}\Bigg(K_0\ R\, D^\mu V^\alpha g_{\alpha\beta} D_{\mu}V^\beta+K_1\ R\, V^\alpha \left(\mathbb{S}^{\mu\nu}\mathbb{S}_{\mu\nu}\right)_{\alpha\beta}V^\beta\\
    &+K_2\ R_{\mu\nu}\, V^\alpha\left(\mathbb{S}^{\mu\lambda}\mathbb{S}_{\lambda}^{\ \nu}\right)_{\alpha\beta}V^\beta+K_3\ R_{\mu\nu\rho\sigma}\, V^\alpha\left(\mathbb{S}^{\mu\nu}\mathbb{S}^{\rho\sigma}\right)_{\alpha\beta}V^\beta\\
    &+K_4\ R_{\mu\nu\rho\sigma}\, D^{\nu}V^\alpha\left(\mathbb{S}^{\mu\lambda}\mathbb{S}_{\lambda}^{\ \sigma}\right)_{\alpha\beta}D^{\sigma}V^\beta+K_5\ D^{\nu}D^{\sigma}R_{\mu\nu\rho\sigma}\, V^\alpha\left(\mathbb{S}^{\mu\lambda}\mathbb{S}_{\lambda}^{\ \sigma}\right)_{\alpha\beta}V^\beta \Bigg)\ .
    \end{aligned}
\end{equation}
It can be shown that every other terms\footnote{This includes higher curvature and higher derivative terms.} not included in Eq.~\eqref{eq:NonMinimalAction} are either $O(\kappa^2)$, local contributions to the metric, or related to the terms above by symmetry properties of the spin operator or the curvature tensor. 
In order to extract  from Eq.~\eqref{eq:NonMinimalAction} a vertex that survives and does not diverge in the classical limit, we must assign the correct dependence on $\hbar$ to the paramenters. While $K_1, K_2$ and $K_3$ are dimensionless parameters, $K_0, K_4$ and $K_5$ have a mass dimension $M^{-2}$, and we observe that  combining $G$ and $m$, there are only two ways to build such terms with a positive integer power of the gravitational coupling, namely
\begin{equation}
   \left[1/m^2\right] = M^{-2}\qquad \text{and} \qquad \left[\frac{1}{\hbar^2}(G m)^{\frac{2}{d-2}}\right] = M^{-2}\ . 
\end{equation}
In particular, the second term leads to an integer power of $G$  only for $d=3,4$. 

By computing the vertex, one observes that for the term proportional to $K_0$, the leading contribution in the transferred momentum  is $O(q^2)$,  while for the one proportional to $K_5$ it is $O(q^4S^2)$. According to Eq.~\eqref{eq:KMOC}, in order to get a classical contribution from both terms we have to compensate for the extra power of $\hbar$, and this implies that both coefficients must be proportional to $(G m)^{\frac{2}{d-2}}$. On the other hand, the leading term proportional to $K_4$ is $O(q^2 S^2)$, and therefore this parameter must be proportional to $1/m^2$. To summarize, 
we can redefine the free parameters in the action in terms of dimensionless quantities as
\begin{equation}
    \begin{gathered}
        K_0=\frac{1}{2}\Omega_1\left(G m\right)^{\frac{2}{d-2}}\ ,\qquad K_1=-\frac{1}{4}C_1\ , \qquad K_2=-\frac{1}{2}C_2\ ,\\
        K_3=\frac{1-H_1}{8}\ ,\qquad K_4=\frac{H_2}{2m^2}\ ,\qquad K_5=\Omega_2\left(G m\right)^{\frac{2}{d-2}}\ .
    \end{gathered}
\end{equation}

Explicitly, expanding~\eqref{eq:NonMinimalAction} at $O(\kappa)$, we obtain the dressed vertex associated to the non-minimal effective action %
\begin{equation}\label{eq:GenericTreeQuadVertex}
    \begin{aligned}
        \hat{\tau}_{V^2h,\text{non-min}}^{\mu\nu}(q)=-\frac{i\, \kappa}{2}\Bigg(&
    -(H_1-1)\, q_{\rho}q_{\sigma}S^{\mu\rho}S^{\nu\sigma}+H_2\, \delta_0^{\mu}\delta_0^{\nu}q_{\rho}q_{\sigma}S^{\rho\lambda}S_{\lambda}^{\ \sigma}+C_1\, S^{\rho\sigma}S_{\rho \sigma}q^{\mu}q^{\nu}\\
    &+C_2\Big(\eta^{\mu\nu}q_{\rho}q_{\sigma}S^{\rho\lambda}S^{\sigma}{}_{\lambda}-q^\lambda\big(q^{\mu}S_{\lambda \sigma}S^{\nu \sigma}+q^{\nu}S_{\lambda \sigma}S^{\mu \sigma}\big)\Big)\Bigg)\ ,
    \end{aligned}
\end{equation}
as well as the ``higher-loop'' dressed vertex
\begin{equation}\label{eq:CounterTermVertex}
    \includegraphics[valign=c,width=0.20\textwidth]{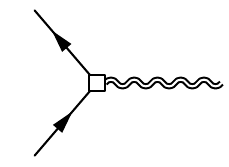}=\hat{\tau}_{V^2h, \text{HL}}^{\mu\nu}(q)=i\, \kappa (G\, m)^{\frac{2}{d-2}}\Bigg(\Omega_1\, m^2\, q^{\mu}q^{\nu}+\Omega_2\, q^{\mu}q^{\nu} q_{\rho}q_{\sigma}S^{\rho\lambda}S_{\lambda}^{\ \sigma}\Bigg)\ .
\end{equation}
In particular, in $d=3$ a tree-level insertion of $ \hat{\tau}_{\Phi^2h, \text{HL}}^{\mu\nu}$ contributes to 3PM, while in $d=4$ it contributes starting from 2PM.
Notice that the $\Omega_1$ piece in Eq.~\eqref{eq:CounterTermVertex} exactly corresponds to the counter-term defined in \cite{Mougiakakos:2020laz}.

Considering both the minimal and non-minimal action, the dressed vertex 
\begin{equation}\label{eq:FinalQuadVertex}
\begin{aligned}
&\hat{\tau}_{V^2h}^{\mu\nu}(q) = \hat{\tau}_{V^2h,\text{min}}^{\mu\nu}(q) + \hat{\tau}_{V^2h,\text{non-min}}^{\mu\nu}(q)\\
&=-\frac{i\, \kappa}{2}\Bigg(2 m^2 \delta^\mu_0 \delta^\nu_0- i\, m\, q_\lambda \left(S^{\mu\lambda}\delta^{\nu}_0+S^{\nu\lambda}\delta_0^{\mu}\right)-H_1q_\lambda q_\sigma S^{\mu\lambda}S^{\nu\sigma}+H_2\, \delta_0^{\mu}\delta_0^{\nu}q_{\rho}q_{\sigma}S^{\rho\lambda}S_{\lambda}^{\ \sigma}\\
&+C_1\, S^{\rho\sigma}S_{\rho \sigma}q^{\mu}q^{\nu}+C_2\Big(\eta^{\mu\nu}q_{\rho}q_{\sigma}S^{\rho\lambda}S^{\sigma}{}_{\lambda}-q^\lambda\big(q^{\mu}S_{\lambda \sigma}S^{\nu \sigma}+q^{\nu}S_{\lambda \sigma}S^{\mu \sigma}\big)\Big)\Bigg)
\end{aligned}
\end{equation}
is associated to the stress-energy tensor of the matter source, and describes the most general stationary rotating matter distribution at second order in the angular momentum, which is spherically symmetric in the non-rotating limit.  Since we have no free parameters in the scalar and dipole terms (other than the mass and angular momentum), such dressed vertex shows that at order $O(S)$ the geometry is uniquely defined, as we commented in subsection~\ref{sec:ExampleSpin1/2}. Moreover, the dimensionless coefficients are normalized in such a way that we recover the minimal dressed vertex in Eq.~\eqref{eq:MinimalVertex} by setting 
\vspace{-0.4pt}
\begin{equation}\label{eq:MinimalLimit}
    H_1=1\ ,\quad H_2=0\ ,\quad C_1=0\ ,\quad C_2=0\ .
\end{equation} 
\vspace{-0.4pt}
Hereafter, we will refer to Eq.~\eqref{eq:MinimalLimit} as the minimal limit. As we will show in detail, while the $C_i$ coefficients are a gauge artifact, the $H_i$ coefficients are physical, in the sense that they determine the multipole structure of the source. 
On the other hand, the higher-loop dressed vertex in Eq.~\eqref{eq:CounterTermVertex}  does not correspond to the stress-energy tensor $T_{\mu\nu}^{(0)}$ of the matter distribution. Indeed, we will see how it contributes only to the renormalization of gauge-dependent singularities in the metric that arise only in $d=3, 4$.

Finally, one can check that both dressed vertices are conserved up to local terms, in the sense that $ q_\mu\hat{\tau}_{V^2h}^{\mu\nu}=O(|\vec q\,| ^2)$ and $ q_\mu\hat{\tau}_{V^2h, \text{HL}}^{\mu\nu}=O(|\vec q\,| ^2)$, as they should be. This fixes the relative coefficients of  the terms proportional to $C_2$ in Eq.~\eqref{eq:GenericTreeQuadVertex}, and does not allow for any other terms. In fact, even neglecting the QFT underlying the dressed vertex, we could have written the stress-energy tensor associated to Eq.~\eqref{eq:GenericTreeQuadVertex} just by constructing the most generic conserved symmetric rank-2 tensor with the objects we have at our disposal. This means that, without considering the field interpretation of the matter source, this procedure can be easily generalized to arbitrary multipoles to build metrics in any dimensions at any order in the PM expansion. In this sense, the results above are universal as far as terms up to quadrupole order are concerned. Considering a field of spin $s$,  we expect that the non-minimal couplings will produce  exactly the same vertex as the spin-1 case at quadrupole level.  In other words, a spin-$s$ field will give rise to the most general vacuum metric generated by a rotating object  with multipoles up to $2s$ order, and with all the higher multipoles vanishing.

\subsection{Loop amplitudes}

\begin{figure}[h]
\centering
\includegraphics[width=0.3\textwidth, valign=c]{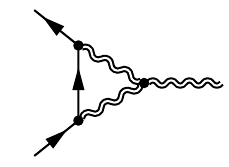}
\caption{1-loop Feynman diagram that contributes to $T_{\mu\nu}^{(1)}(q)$.}
\label{fig:1_Loop}
\end{figure}

\begin{figure}[h]
\centering
\includegraphics[width=0.3\textwidth, valign=c]{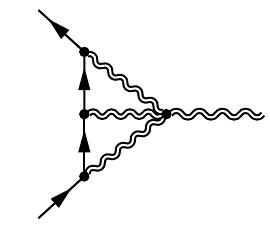}
\includegraphics[width=0.3\textwidth, valign=c]{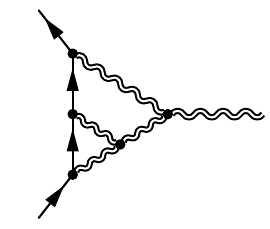}
\caption{2-loop Feynman diagrams that contribute to $T_{\mu\nu}^{(2)}(q)$.}
\label{fig:2_Loop}
\end{figure}

As already mentioned, the dressed vertex in Eq.~\eqref{eq:GenericTreeQuadVertex} is directly related to the stress-energy tensor $T_{\mu\nu}^{(0)}(q)$ of the matter source, which through Eq.~\eqref{eq:MetricFromEMT} leads to the metric at 1PM. In order to get higher order terms, one has to compute self-interaction contributions to the stress-energy tensor, which are produced by the loop amplitudes in Eq.~\eqref{eq:EMTfromAMPpic}. At 1-loop level the only diagram that contributes is the one in Fig.~\ref{fig:1_Loop},
while at  2-loop level the diagrams that contribute to $T_{\mu\nu}^{(2)}(q)$ are shown in Fig.~\ref{fig:2_Loop}. The computation is performed following the procedure outlined in subsection~\ref{sec:ClassicalSpin}.
\begin{figure}[h]
\centering
\includegraphics[width=0.3\textwidth, valign=c]{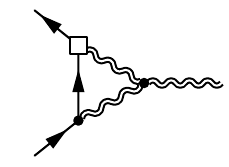}
\caption{1-loop insertion of the higher-loop vertex that renormalizes the 3PM metric in $d=4$.}
\label{fig:1LoopBox}
\end{figure}
The loop amplitudes exhibit infrared divergences for specific values of $d$, and although in general the computation of $h_{\mu\nu}^{(n)} (x)$ by means of the Fourier transform in Eq.~\eqref{eq:MetricFromEMT} smears them, they lead to singularities in the metric  in $d=3$ and $d= 4$. These divergences have to be renormalized by the insertion of diagrams involving the higher-loop vertex  in Eq.~\eqref{eq:CounterTermVertex} used as counter-terms, depending on the specific value of $d$ as explained in the last subsection. At 2PM the only dimension in which a singularity appears in the metric is  $d=4$, and the singularity is cured by the tree-level insertion of $\hat{\tau}_{V^2h, \text{HL}}^{\mu\nu}$. In this case, we can redefine the coefficients in~\eqref{eq:CounterTermVertex} as 
\begin{equation}
    \begin{aligned}
        \Omega_1\big|_{d=4}&=\frac{\Omega_1^{\mathrm{renorm.}}}{d-4}+\Omega_1^{\mathrm{free}}\ ,\\
        \Omega_2\big|_{d=4}&=\frac{\Omega_2^{\mathrm{renorm.}}}{d-4}+\Omega_2^{\mathrm{free}}\ ,
    \end{aligned}
\end{equation}
where $\Omega_i^{\mathrm{renorm.}}$ is the term that has to be fixed to cancel the poles, and $\Omega_i^{\mathrm{free}}$ is the remaining free parameter. Then, within our conventions we have to fix
\begin{equation}\label{eq:RenormalizationChoice}
    \Omega_1^{\mathrm{renorm.}}=\frac{1}{9\pi} \qquad \mathrm{and} \qquad \Omega_{2}^{\mathrm{renorm.}}=\frac{H_1+2H_2-1}{60\pi}\ ,
\end{equation}
such that the metric is now finite in $d=4$ and some logarithmic terms arise in the radial dependence of the metric. 

At 2-loop level,  the metric is divergent in both $d=3$ and $d=4$. 
As we already discussed, in $d=4$ the 3PM metric is renormalized by a 1-loop insertion of the higher-loop vertex,  as in Fig.~\ref{fig:1LoopBox}. 
Consistently, the parameters defined in Eq.~\eqref{eq:RenormalizationChoice} renormalize the 3PM metric in $d=4$, and the final result is finite. 
In the $d=3$ case,  the metric is renormalized by a tree-level insertion of the higher-loop vertex. However, even if the same renormalization procedure that we carried out for $d=4$ can be pursued, obtaining a result for every value of the gauge parameter $\alpha$, we  notice that in $d=3$ the harmonic gauge ($\alpha=1$) naturally eliminates every poles, and the metric becomes finite\footnote{In this case the tree-level insertion of the higher-loop vertex only adds two gauge redundancies to the metric.}. We checked this explicitly up to 2-loops, and at the best of our knowledge there is no principle for which such gauge guaranties finiteness at every PM and multipole orders. In fact in $d=4$ at 1-loop, one can define a particular gauge choice (choosing a specific value of $\alpha$) in which $h_{\mu\nu}^{(2, 0)}(r)$ is finite, but $h_{\mu\nu}^{(2, 2)}(r)$ is not. 

From the point of view of the Einstein equations, the divergences discussed above are interpreted as a consequence of the fact that, for a specific  gauge choice, imposing that the metric is a series expansion in powers of $\rho (r)$ is not consistent. Similarly, the presence of the free parameters $\Omega_i^{\mathrm{free}}$ is interpreted  as a redundancy of the gauge defined in Eq.~\eqref{eq:GaugeCondition}.
In the next section we will discuss in detail the properties of the metric and give its full expression  at 1PM. Although we have also managed to perform such calculation up to 2-loops and for an arbitrary value of $\alpha$, higher orders in the PM expansion do not give any more physical information than the leading order, and since the full expressions are quite involved, we only provide the analytic form of $h_{\mu\nu}^{(2)}(r)$ and $h_{\mu\nu}^{(3)}(r)$ in   an auxiliary Mathematica file \cite{AnchillaryFiles}.

\section{Rotating metrics in any dimensions}\label{sec:rotmetric}

From Eq.~\eqref{eq:FinalQuadVertex}, by considering the Fourier transform in Eq.~\eqref{eq:MetricFromEMT}, we can compute the metric induced by the most generic rotating matter distribution (which is spherically symmetric in the non-rotating case) at quadrupole order at 1PM. At tree-level we have already computed the metric up to $O(S)$ in Sec.~\ref{sec:ExampleSpin1/2}, so we just need to determine the quadrupole part, which reads
\begin{equation}\label{eq:QuadMetric}
    \begin{aligned}
        h_{00}^{(1, 2)}(r)&=\frac{2(d-2)\Big(H_2(d-2)+H_1\Big)}{d-1}\frac{r^2S_{k_1k_2}S^{k_1k_2}-d\, x^{k_1}x^{k_2}S_{k_1}{}^{k_3}S_{k_2k_3}}{mr^4}G \rho(r)\ ,\\
        h_{0i}^{(1, 2)}(r)&=0\ ,\\
        h_{ij}^{(1, 2)}(r)&=-\frac{2(d-2)}{(d-1)mr^4}\Bigg(-C_1(d-1)d\, x_ix_jS_{k_1k_2}S^{k_1k_2}-r^2(d-1)\Big(2C_2+H_1\Big)S_{ik}S_{j}{}^{k}\\
        &+r^2\Big(C_1(d-1)+H_1-H_2\Big)S_{k_1k_2}S^{k_1k_2}\delta_{ij}+d\, C_2(d-1)x^{k_1}S_{k_1k_2}\Big(x_jS_{i}{}^{k_2}+x_iS_{j}{}^{k_2}\Big)\\
        &+d\, x^{k_1}x^{k_2}\Big((d-1)H_1S_{ik_1}S_{jk_2}+(H_2-H_1)S_{k_1}{}^{k_3}S_{k_2k_3}\delta_{ij}\Big)\Bigg)G\rho(r)\ . 
    \end{aligned}
\end{equation}
We notice that, at this level, the metric does not depend on the gauge parameter $\alpha$ introduced in Eq.~\eqref{eq:GaugeCondition} since the differences between the de Donder and the harmonic gauge start to appear at 2PM in the expression of the self-interacting graviton vertices.

\subsection{Eliminating redundant parameters}\label{sec:EliminateGauge}

In the last subsection we recovered the metric induced by a rotating matter source at quadrupole order, which depends on four free parameters. However, it can be shown that some of these  are gauge artifacts, which can be eliminated by an infinitesimal coordinate transformation.

Let us restrict to the case of $\alpha=1$ (harmonic gauge), even though the same argument can be generalized. Consider a generic infinitesimal coordinate transformation $x'_\mu=x_\mu+\xi_\mu(x)$, such that the metric perturbation in the new frame reads
\begin{equation}\label{eq:MetricPrime}
    h'_{\mu\nu}=h_{\mu\nu}-(\partial_\mu\xi_\nu+\partial_\nu\xi_\mu)\ .
\end{equation}
By definition, in the harmonic gauge
\begin{equation}
    \Box x^{\mu}=0\ ,
\end{equation}
and if we want to make a coordinate transformation preserving this gauge, we just have to impose
\begin{equation}
    \Box x'^{\mu}=0\quad \rightarrow\quad \Box \xi^\mu=0\ .
\end{equation}
Moreover, in order to preserve stationarity the infinitesimal shift $\xi^\mu$ must be time-independent and in addition we choose it such that $\xi^0=0$, so that only the spatial components of the metric are transformed.
With these choices, since in arbitrary dimension 
\begin{equation}
    \Box \rho(r)=0\ ,
\end{equation}
the most generic harmonic shift is 
\begin{equation}
    \xi^i=\sum_{\ell=0}^{+\infty} \mathcal{T}^{i,A_\ell}\partial_{A_\ell}\rho(r)\ ,
\end{equation}
where $\mathcal{T}$ is a generic constant tensor, and we will use hereafter the shorthand notation $A_\ell=a_1\cdots a_\ell$, with $\partial_{A_\ell}=\partial_{a_1}\cdots\partial_{a_\ell}$.
Finally, since $\xi^i$ must have the dimensions of a length, at a certain order in the derivatives of the harmonic function we have to define a dimensionful quantity that compensates for the extra length dimensions.

The most generic harmonic function linear in $G$ that parameterizes a coordinate transformation at quadrupole order inside the gauge reads
\begin{equation}\label{eq:epsilonC1C2}
  \xi^i=\frac{G}{m}\Big(A\, S^{ik}S_{k}{}^{j}+B\, S^{lm}S_{lm}\delta^{ij}\Big)\partial_j\rho(r)  \ .
\end{equation}
Choosing the dimensionless coefficients as
\begin{equation}
    A=2C_2 \qquad \text{and} \qquad B=C_1\ 
\end{equation}
and computing Eq.~\eqref{eq:MetricPrime} we will end up with a metric independent of $C_1$ and $C_2$. This  means that such coefficients in Eq.~\eqref{eq:QuadMetric} are only a gauge artifact, while $H_1$ and $H_2$ are physical because there is no transformation inside the gauge that can cancel them.

There are two additional shifts that can be defined at higher PM order up to quadrupole terms. In particular, at order $O(S^0)$ one can consider
\begin{equation}\label{eq:epsilon1}
    \xi^i_1=(Gm)^{\frac{d}{d-2}}\, \tilde{\Omega}_1\,  \partial^i\rho(r)\ ,
\end{equation}
while at order $O(S^2)$ one gets
\begin{equation}\label{eq:epsilon2}
    \xi^i_2=\frac{1}{m^2}(Gm)^{\frac{d}{d-2}}\, \tilde{\Omega}_2\, S_{l}{}^kS_{km}\partial^i\partial^l\partial^m\rho(r)\ .
\end{equation}
From Eqs.~\eqref{eq:epsilon1} and~\eqref{eq:epsilon2} one can repeat the previous argument and show that the coefficients $\Omega_1$ and $\Omega_2$ in Eq.~\eqref{eq:CounterTermVertex} are a gauge artifact. 

In summary, besides the spacetime mass $m$ and spin tensor $S^{\mu\nu}$, the solution depends on \emph{two} further physical parameters, namely $H_1$ and $H_2$, while every other free coefficient parameterizes a gauge redundancy. This means that in the non-minimal action in Eq.~\eqref{eq:NonMinimalAction} every term can be reabsorbed by a coordinate transformation except for the terms proportional to $K_3$ and $K_4$, which are physical and are not associated with a gauge transformation.
However, in $d=3$ this is not the end of the story. Indeed, such case is special since we can write $S^{ij}=\epsilon^{ijk}S_k$, where $S^{k}$ is the angular momentum vector and $\epsilon^{ijk}$ is the Levi-Civita symbol.  Replacing this relation inside the metric in $d=3$ and fixing
\begin{equation}
    A=H_2-H_1 \qquad \text{and} \qquad B=-\frac{1}{2}H_1\ ,
\end{equation}
we find that the entire metric depends on $H_1$ and $H_2$ only through the combination $H_1+H_2$,
showing that the $D=4$ metric effectively depends only on one extra parameter. The physical interpretation of $H_1$ and $H_2$, as well as their degeneracy in four dimensions, will be discussed in the next subsection.

\subsection{Multipole moments}\label{sec:multipolemoments}
Having identified the physical parameters of the solution, we now turn our attention to their physical interpretation. 
The metric in Eq.~\eqref{eq:QuadMetric}, in which the coefficients $H_1$ and $H_2$ enter, is truncated at the leading quadrupole order in a multipole expansion, and so it is natural to think that such coefficients are associated with the quadrupole moment(s) of the source. 

The concept of multipole expansion in General Relativity has a long history, pioneered by Geroch and Hansen~\cite{Geroch:1970cc, Geroch:1970cd, Hansen:1974zz}, who defined a multipole expansion in a gauge invariant framework. Later, Thorne introduced the so-called Asymptotically Cartesian Mass Centered~(ACMC) coordinates, a particular reference frame in which it is possible to unambiguously extract the multipole moments directly from the asymptotic behavior of the metric~\cite{Thorne:1980ru}. Albeit not gauge invariant, Thorne's formalism is more intuitive, and it can be shown that the two definitions of multipole moments coincide~\cite{Gursel1983}. While most of the knowledge on this topic is limited to the $D=4$ case (see Refs.~\cite{Cardoso:2016ryw,Mayerson:2022ekj} for a review), there have been attempts to generalize the multipole expansion formalism to $D=5$ in the Geroch-Hansen framework~\cite{Tanabe:2010ax} and using the ACMC coordinates \`a la Thorne~\cite{Heynen:2023sin}. 

Let us discuss the original formalism of Thorne~\cite{Thorne:1980ru}, which is also used in~\cite{Heynen:2023sin}. Considering the linearized metric
\begin{equation}
    g_{\mu\nu}=\eta_{\mu\nu}+\kappa h^{(1)}_{\mu\nu}+...\ ,
\end{equation}
which is nothing but Eq.~\eqref{eq:expansioninh} truncated at first order, and defining the trace-reversed perturbation as
\begin{equation}\label{eq:gammaDefinition}
    \gamma_{\mu\nu}= \kappa h^{(1)}_{\mu\nu}-\frac{\kappa}{2}\eta_{\mu\nu}h^{(1)}\ ,
\end{equation}
the harmonic gauge condition and the Einstein equations respectively are
\begin{equation}\label{eq:HarmonicGaugeTR}
    \partial^\mu\gamma_{\mu\nu}=0 \qquad \text{and} \qquad \Box \gamma_{\mu\nu}=0\ .
\end{equation}
As we already noticed in subsection~\ref{sec:EliminateGauge}, we can build harmonic functions in arbitrary dimensions at any order by means of derivatives of the harmonic function $\rho(r)$. In this framework, the most generic solution of the linearized vacuum Einstein equations can be written as
\begin{equation}\label{eq:gammaMetric}
    \begin{aligned}
    \gamma_{00}&=\sum_{\ell=0}^{+\infty}\mathcal{M}_{A_\ell}\partial_{A_\ell}\rho(r)\ , \\
    \gamma_{0i}&=\sum_{\ell=0}^{+\infty}\mathcal{J}_{i, A_{\ell}}\partial_{A_\ell}\rho(r)\ ,\\
    \gamma_{ij}&=\sum_{\ell=0}^{+\infty}\mathcal{G}_{ij, A_{\ell}}\partial_{A_\ell}\rho(r)\ ,
    \end{aligned}
\end{equation}
where $\mathcal{M}_{A_\ell}$, $\mathcal{J}_{i, A_{\ell}}$ and $\mathcal{G}_{ij, A_{\ell}}$ are generic constant tensors which are completely symmetric and traceless in the indices $A_\ell$, and following~\cite{Thorne:1980ru, Heynen:2023sin} we will denote these indices as $\{A_\ell\}_{\text{STF}}$ (symmetric and trace-free); moreover, $\mathcal{G}_{ij, A_{\ell}}$ is also symmetric with respect to $ij$ but not traceless, so in a shorthand notation we express all these symmetries by writing $\mathcal{G}_{(ij), \{A_\ell\}_{\text{STF}}}$. In particular, in $d$ spatial dimensions, these multipole tensors are $SO(d)$ tensors, and they can be decomposed in irreducible representations of such rotation group.
For the moment, let us restrict to the quadrupole case ($\ell=2$) and  $d=3$ spatial dimensions. 
We can rewrite the trace-reversed perturbation in terms of irreducible representations as (see Appendix~\ref{app:multipole}) 
\begin{equation} \label{eq:gammaMetricquadrupole}
\begin{aligned}
   & \gamma_{00}\big|_{\ell=2}^{d=3}=\mathcal{M}_{\{a_1a_2\}_{\text{STF}}}\partial_{a_1}\partial_{a_2}\left(\frac{1}{r}\right)\ ,\\
   & \gamma_{0i}\big|_{\ell=2}^{d=3}=\mathcal{J}^{(1)}_{a_1}\partial_{a_1}\partial_{i}\left(\frac{1}{r}\right)+\mathcal{J}^{(2)}_{\{ia_1a_2\}_{\text{STF}}}\partial_{a_1}\partial_{a_2}\left(\frac{1}{r}\right)+\epsilon_{ia_1a_2}\mathcal{J}^{(3)}_{\{a_1a_3\}_{\text{STF}}}\partial_{a_2}\partial_{a_3}\left(\frac{1}{r}\right)\ ,\\
   &  \gamma_{ij}\big|_{\ell=2}^{d=3}=\delta_{ij}\mathcal{G}^{(1)}_{\{a_1a_2\}_{\text{STF}}}\partial_{a_1}\partial_{a_2}\left(\frac{1}{r}\right)+\mathcal{G}^{(2)}\partial_{i}\partial_{j}\left(\frac{1}{r}\right)+\mathcal{G}^{(3)}_{\{(i|a_1\}_{\text{STF}}}\partial_{|j)}\partial_{a_1}\left(\frac{1}{r}\right)\\
     &+\mathcal{G}^{(4)}_{\{ija_1a_2\}_{\text{STF}}}\partial_{a_1}\partial_{a_2}\left(\frac{1}{r}\right)+\epsilon_{(i|a_1a_2}\mathcal{G}^{(5)}_{a_1}\partial_{a_2}\partial_{|j)}\left(\frac{1}{r}\right)+\epsilon_{(i|a_1a_2}\mathcal{G}^{(6)}_{\{a_1a_3|j)\}_{\text{STF}}}\partial_{a_2}\partial_{a_3}\left(\frac{1}{r}\right)\ ,
\end{aligned}
\end{equation}
where all the tensors are constant and the superscripts are just labelling the different tensors. Now imposing the harmonic gauge condition in Eq.~\eqref{eq:HarmonicGaugeTR} one has to fix 
\begin{equation}
    \mathcal{J}^{(2)}=0\ , \qquad \mathcal{G}^{(1)}=-\frac{1}{2}\mathcal{G}^{(3)}\ ,\qquad \mathcal{G}^{(4)}=0\ ,\qquad \mathcal{G}^{(6)}=0\ . 
\end{equation}
Moreover, as we did in the last subsection, we can define a coordinate transformation inside the harmonic gauge where
\begin{equation}
    \gamma'_{\mu\nu}=\gamma_{\mu\nu}-\partial_\mu\xi_\nu-\partial_\nu\xi_\mu+\eta_{\mu\nu}\partial^\alpha\xi_\alpha\ .
\end{equation}
We can exploit this gauge freedom to choose
\begin{equation}
    \begin{gathered}
\xi_0=J^{(1)}_{a_1}\partial_{a_1}\left(\frac{1}{r}\right) \ ,\\
        \xi_i=-\mathcal{G}^{(1)}_{ja_1}\partial_{a_1}\left(\frac{1}{r}\right)+\frac{1}{2}\epsilon_{ja_1a_2}\mathcal{G}^{(5)}_{a_1}\partial_{a_2}\left(\frac{1}{r}\right)+\frac{1}{2}\mathcal{G}^{(2)}\partial_{j}\left(\frac{1}{r}\right)\ ,
    \end{gathered}
\end{equation}
obtaining a trace-reversed perturbation that reads
\begin{equation}\label{eq:gammaijvanishes}
\begin{aligned}
    \gamma_{00}\big|_{\ell=2}^{d=3}&=\mathcal{M}_{\{a_1a_2\}_{\text{STF}}}\partial_{a_1}\partial_{a_2}\left(\frac{1}{r}\right)\ ,\\
    \gamma_{0i}\big|_{\ell=2}^{d=3}&=\epsilon_{ia_1a_2}\mathcal{J}^{(3)}_{\{a_1a_3\}_{\text{STF}}}\partial_{a_2}\partial_{a_3}\left(\frac{1}{r}\right)\ ,\\
     \gamma_{ij}\big|_{\ell=2}^{d=3}&=0\ .
\end{aligned}
\end{equation}
This argument can be generalized to arbitrary multipole orders, exactly as in~\cite{Thorne:1980ru}, and the final outcome is that in $d=3$ there are two independent towers of multipole tensors, namely the mass multipoles $\mathcal{M}_{\{A_\ell\}_{\text{STF}}}$ and the current multipoles $\mathcal{J}^{(3)}_{\{A_\ell\}_{\text{STF}}}$, which are gauge invariant within ACMC transformations. 

Let us now discuss the quadrupolar case ($\ell=2$) but in $d=4$. In this case we have to decompose the perturbation in terms of $SO(4)$ irreducible representations and, besides STF tensors, there will appear two more structures (see Appendix~\ref{app:multipole} for details). Following~\cite{Heynen:2023sin}, we will indicate as ASTF (anti-symmetric and trace-free) a tensor $\mathcal{T}_{\{b_1b_2, A_\ell\}_\text{ASTF}}$, which is STF with respect to $A_\ell$, anti-symmetric on the $b$'s, and trace-free with respect to all indices. Furthermore, irreducible representations of $SO(4)$ lead to another kind of tensors (not recognized in~\cite{Heynen:2023sin}) that we call RSTF (Riemann-symmetric and trace-free), and we indicate them as $\mathcal{T}_{\{ib_1,jb_2,A_\ell\}_\text{RSTF}}$. They have the same symmetries of the Riemann tensor for the first four indices, are STF with respect to $A_\ell$'s, and are trace-free with respect to all indices. We can now express the trace-reversed metric perturbation in terms of irreducible representations as 
\begin{equation} \label{eq:gammaMetricquadrupoleD5}
\begin{aligned}
    \gamma_{00}\big|_{\ell=2}^{d=4}&=\mathcal{M}_{\{a_1a_2\}_{\text{STF}}}\partial_{a_1}\partial_{a_2}\left(\frac{1}{\pi r^2}\right)\ ,\\
    \gamma_{0i}\big|_{\ell=2}^{d=4}&=\mathcal{J}^{(1)}_{a_1}\partial_{a_1}\partial_{i}\left(\frac{1}{\pi r^2}\right)+\mathcal{J}^{(2)}_{\{ia_1a_2\}_{\text{STF}}}\partial_{a_1}\partial_{a_2}\left(\frac{1}{\pi r^2}\right)+\epsilon_{ib_1b_2a_1}\mathcal{J}^{(3)}_{\{b_1b_2,a_2\}_{\text{ASTF}}}\partial_{a_1}\partial_{a_2}\left(\frac{1}{\pi r^2}\right)\ ,\\
     \gamma_{ij}\big|_{\ell=2}^{d=4}&=\delta_{ij}\mathcal{G}^{(1)}_{\{a_1a_2\}_{\text{STF}}}\partial_{a_1}\partial_{a_2}\left(\frac{1}{\pi r^2}\right)+\mathcal{G}^{(2)}\partial_{i}\partial_{j}\left(\frac{1}{\pi r^2}\right)+\mathcal{G}^{(3)}_{\{(i|a_1\}_{\text{STF}}}\partial_{|j)}\partial_{a_1}\left(\frac{1}{\pi r^2}\right)\\
     &+\mathcal{G}^{(4)}_{\{ija_1a_2\}_{\text{STF}}}\partial_{a_1}\partial_{a_2}\left(\frac{1}{\pi r^2}\right)+\epsilon_{(i|b_1b_2a_1}\mathcal{G}^{(5)}_{\{b_1b_2\}_\text{ASTF}}\partial_{a_1}\partial_{|j)}\left(\frac{1}{\pi r^2}\right)\\
    &+\epsilon_{(i|b_1b_2a_1}\mathcal{G}^{(6)}_{\{b_1b_2,a_2|j)\}_{\text{ASTF}}}\partial_{a_1}\partial_{a_2}\left(\frac{1}{\pi r^2}\right)+\mathcal{G}^{(7)}_{\{ib_1,jb_2\}_\text{RSTF}}\partial_{b_1}\partial_{b_2}\left(\frac{1}{\pi r^2}\right)\ .
\end{aligned}
\end{equation}
Notice that $\mathcal{G}^{(7)}$ is not manifestly symmetric in $ij$ since it has a Riemann-like symmetry; however, once symmetrized in $b_1b_2$ (due to the contraction with the derivatives), it becomes symmetric also with respect to $ij$.

Comparing this result with the work in~\cite{Heynen:2023sin}, one can notice that in Eq.~\eqref{eq:gammaMetricquadrupoleD5} there is an extra structure, namely $\mathcal{G}^{(7)}$. The existence of such new tensor is very clear from a group theory point of view (see Appendix~\ref{app:multipole}), in which by counting the number of degrees of freedom one gets
\begin{equation}
    \#\Bigg[\mathcal{G}^{(7)}_{\{ib_1,jb_2\}_\text{RSTF}}+\mathcal{G}^{(6)}_{\{b_1b_2,a_1a_2\}_{\text{ASTF}}}\Bigg]=40\ ,
\end{equation}
conversely to what is discussed in~\cite{Heynen:2023sin}, in which 40 components are assigned to $\mathcal{G}^{(6)}$ only. 

As done previously for the $d=3$ case, after imposing the harmonic gauge condition, we can find a suitable coordinate transformation inside the gauge such that the perturbation finally reads
\begin{equation} \label{eq:gammaMetricquadrupoleD5final}
\begin{aligned}
    \gamma_{00}\big|_{\ell=2}^{d=4}&=\mathcal{M}_{\{a_1a_2\}_{\text{STF}}}\partial_{a_1}\partial_{a_2}\left(\frac{1}{\pi r^2}\right)\ ,\\
    \gamma_{0i}\big|_{\ell=2}^{d=4}&=\epsilon_{ib_1b_2a_1}\mathcal{J}^{(3)}_{\{b_1b_2,a_2 \}_{\text{ASTF}}}\partial_{a_1}\partial_{a_2}\left(\frac{1}{\pi r^2}\right)\ ,\\
     \gamma_{ij}\big|_{\ell=2}^{d=4}&=\mathcal{G}^{(7)}_{\{ib_1,jb_2\}_\text{RSTF}}\partial_{b_1}\partial_{b_2}\left(\frac{1}{\pi r^2}\right)\ .
\end{aligned}
\end{equation}
Since now $\gamma_{ij}\neq 0$, we have an extra degree of freedom at quadrupole order with respect to the $d=3$ case. This argument can be generalized to arbitrary dimensions and to any multipole order, showing the existence of a new tower of independent multipole tensors $\mathcal{G}^{(7)}_{\{ib_1,jb_2,A_\ell\}_\text{RSTF}}$ that we call \textit{stress multipoles}\footnote{The name is inspired by the fact that $g_{ij}$ is induced by $T_{ij}$, which is the stress-part of the matter source.}. To summarize, Eq.~\eqref{eq:gammaMetricquadrupoleD5final} shows that at each multipole order the spacetime is characterized by \emph{three} independent multipole tensors, which however in the case of $D=4$ reduce to two only, due to group theory properties of $SO(3)$ and the specific form of the Einstein equations.

Inspired by the above argument and by the metric in Eq.~\eqref{eq:QuadMetric}, we conjecture the possibility of defining an ACMC coordinate system in arbitrary dimensions and the existence of a new independent multipole tensor associated with the spatial part of the metric. To this end we give the generic expression of a stationary metric in arbitrary dimensions expanded in a multipole series as
\begin{equation}\label{eq:GenericMultipoleMetric}
\begin{aligned}
    g_{00}&=1-\frac{4(d-2)}{d-1}Gm\rho(r)+\sum_{\ell=1}^{+\infty}\frac{2(d-2)}{d-1}\frac{G m \rho(r)}{r^\ell}\mathbb{M}^{(\ell)}_{A_\ell}N_{A_\ell}+\cdots\ ,\\
    g_{0i}&=-2(d-2)\frac{Gm\rho(r)}{r}\left(\frac{1}{m}S^{ik}\frac{x_k}{r}\right)+2(d-2)\sum_{\ell=2}^{+\infty}\frac{Gm\rho(r)}{r^\ell}\mathbb{J}^{(\ell)}_{i, A_\ell}N_{A_\ell}+\cdots\ ,\\ 
    g_{ij}&=-\delta_{ij}-\frac{4}{d-1}Gm\rho(r)\delta_{ij}+\sum_{\ell=1}^{+\infty}\frac{2(d-2)}{d-1}\frac{G m \rho(r)}{r^\ell}\tilde{\mathbb{G}}^{(\ell)}_{ij,A_\ell}N_{A_\ell}+\cdots\ ,
\end{aligned}
\end{equation}
with $N_{A_\ell}=\frac{x_{a_1}\cdots\, x_{a_\ell}}{r^{\ell}}$ and where the ellipses stand for non-gauge invariant contributions.
Indeed, in order to assure gauge invariance, it is important to define the quadrupole tensors up to terms $\delta_{a_m a_n}, \delta_{i a_m}$ or $\delta_{j a_m}$. Loosely speaking, in Eq.~\eqref{eq:GenericMultipoleMetric} we are neglecting terms in which $N_{A_\ell}$'s contract some $\delta$'s.
Therefore, within a ``generalized'' ACMC coordinate transformation in arbitrary dimensions, the expression written in Eq.~\eqref{eq:GenericMultipoleMetric} is conjectured to be invariant.
$\mathbb{M}^{(\ell)}_{A_\ell}$ are the standard mass multipoles, $\mathbb{J}^{(\ell)}_{A_\ell}$ correspond to the so-called current multipoles, and $\tilde{\mathbb{G}}^{(\ell)}_{ij,A_\ell}$ contain the new multipole tensors that we have just discovered, whose expression is
\begin{equation}
    \mathbb{G}_{ij, A_\ell}^{(\ell)}=\tilde{\mathbb{G}}_{ij, A_\ell}^{(\ell)}+\frac{1}{2}\delta_{ij}\Big(\mathbb{M}^{(\ell)}_{A_\ell}-\tilde{\mathbb{G}}_{kk, A_\ell}^{(\ell)}\Big)\ ,
\end{equation}
as can be easily derived from the definition~\eqref{eq:gammaDefinition}.
Since the latter multipoles are associated with the spatial part of the metric, it is natural to dub them \emph{stress} multipole moments. 

It is important to notice that in Eq.~\eqref{eq:GenericMultipoleMetric} the explicit dependence on $G$ is included for dimensional reasons, and that in general the multipole tensors themselves can have an explicit dependence on the Newton constant. The formalism described above is indeed completely general, and takes into account also the possibility of having \emph{intrinsic}, \textit{i.e.} not necessarily spin-induced, multipoles\footnote{This is the case, for example, of a static body deformed away from spherical symmetry~\cite{Raposo:2018xkf,Raposo:2020yjy}.}. In that case, in addition to the mass and the angular momenta, a new length scale is present, and it is always possible to rewrite it in units of the fundamental scale $Gm$, thus introducing extra powers of $G$ in the multipolar expansion (an example of this case is the black ring discussed in Sec.~\ref{sec:BlackRing}). 
In order to consider such intrinsic multipoles in a bottom-up approach, from the amplitude perspective one can consider a non-minimal action that, instead of terms quadratic in the spin tensor like in Eq.~\eqref{eq:NonMinimalAction}, contains new independent tensors for each term, defined similarly to Eq.~\eqref{eq:DefinitionSpinTensor}.

Let us now restrict to spin-induced multipole moments. In the specific case in which the matter source is axis-symmetric\footnote{In $D=4$ this corresponds to the usual symmetry along the rotational axis, in $D>4$ one can define a similar symmetry in which every rotational axis can be exchanged.}, the mass and the new stress multipoles contain only even orders of the spherical harmonics (even $\ell$), while the current multipoles are formed only by odd orders of them (odd $\ell$). The metric in Eq.~\eqref{eq:QuadMetric} describes the most generic stationary axis-symmetric matter configuration at quadrupole order, and therefore, according to our definitions, we can write the spin-induced mass and stress quadrupoles of a generic rotating source in arbitrary dimensions as
\begin{align}
    \mathbb{M}^{(2)}_{a_1a_2}&=-\frac{1}{m^2}d\Big(H_1+(d-2)H_2\Big)S_{a_1k}S_{a_2}{}^{k}\ , \label{eq:MassQuadrupoleGenericD} \\
    \mathbb{G}^{(2)}_{ij,a_1a_2}&=-\frac{1}{m^2}d(d-1)H_1S_{(i|a_1}S_{|j)a_2}\ .\label{eq:StressQuadrupoleGenericD}
\end{align}
It is now evident that the parameters $H_1$ and $H_2$ of our generic solution are actually related to the mass quadrupole moment $\mathbb{M}^{(2)}_{a_1a_2}$ and to the new stress quadrupole moment $\mathbb{G}^{(2)}_{ij,a_1a_2}$. In the next section we will show that this is indeed the case for the Myers-Perry and black ring solutions, which have a non-vanishing stress quadrupole moment.

In order to see the degeneracy that occurs in $d=3$ in a ``gauge invariant'' way, we just need to rewrite the spin tensor as a spin vector, and the result reads
\begin{align}
         \mathbb{M}^{(2)}_{a_1a_2}N_{a_1a_2}\Big|_{d=3}&=\frac{3}{m^2}(H_1+H_2)(S\cdot x)^2+\cdots\ ,\label{eq:MassQuadrupoleD4} \\
         \tilde{\mathbb{G}}^{(2)}_{ij,a_1a_2}N_{a_1a_2}\Big|_{d=3}&=\delta_{ij}\mathbb{M}^{(2)}_{a_1a_2}N_{a_1a_2}\Big|_{d=3}+\cdots\ ,\label{eq:StressQuadrupoleD4}
\end{align}
where again the ellipses indicate that the expressions are equal up to terms that contain contraction of the coordinates and some $\delta$'s. Note that, only in this $d=3$ case, $\tilde{\mathbb{G}}^{(2)}_{ij,a_1a_2}$ is fixed in terms of $\mathbb{M}^{(2)}_{a_1a_2}$ and therefore $\mathbb{G}^{(2)}_{ij,a_1a_2}=0$. This result generalizes to all orders in the multipole expansion.

\section{Particular cases}\label{sec:particular}

In this section we compare the metric obtained with our amplitude-based approach to the PM expansion of known vacuum solutions in $D\geq4$, and discuss their multipolar structure. We will see that, in $D=4$, in order to reproduce a specific solution at quadrupole order we only need to fix a combination of $H_1$ and $H_2$, while in $D=5$ we have to fix both of them independently, proving explicitly the existence of the stress multipole moment. Furthermore, we define the ``simplest'' metric as the metric associated with a minimally-coupled theory. We put it in relation with specific cases, proving that Kerr black holes are the simplest solution in $D=4$, while Myers-Perry black holes and black rings in $D=5$ are not.

\subsection{Hartle-Thorne metric in $D=4$}\label{sec:HartleThorneComparison}

The Hartle-Thorne metric~\cite{Hartle:1967he, Hartle:1968si} is the metric induced by the most generic rotating matter distribution in $D=4$ up to spin-induced quadrupole order. 
As shown in Appendix~\ref{app:Hartle-Thorne}, we can rewrite the Hartle-Thorne metric in harmonic coordinates in order to compare it with the amplitude result. This transformation introduces some free numerical coefficients due to gauge redundancies that can be mapped to the free parameters of our solution. With such an identification, up to 3PM, we have proved that the Hartle-Thorne metric and the amplitude-based one match exactly. Such comparison has been done in the special frame in which the Hartle-Thorne metric is defined, namely with the angular momentum aligned with the $z$-axis. In order to move to such frame, in terms of the spin tensor we have to choose
\begin{equation}
 S_{ij}=\begin{pNiceMatrix}[columns-width=auto]
        0 & J & 0 \\
        -J & 0 & 0 \\
        0 & 0 & 0
    \end{pNiceMatrix}\ ,
\end{equation}
where $J\equiv a/m$ can be identified with the physical angular momentum of the spacetime through an expansion in the far-field limit.

The temporal component of the Hartle-Thorne metric in harmonic coordinates reads
\begin{equation}
    g_{00}^{\rm HT}=1-\frac{2 G m}{r}+\frac{a^2 G m \zeta}{r^3}\left(3\frac{z^2}{r^2}-1\right) +O(G^2,a^3)\ ,
\end{equation}
where $\zeta$ parametrizes the mass quadrupole\footnote{Due to the axisymmetry, the quadrupole moment tensor is defined by a single parameter.}. By setting 
\begin{equation}\label{eq:HTcondition}
    H_1+H_2=\zeta\ ,
\end{equation}
the amplitude-based metric fully reproduces the Hartle-Thorne solution. Therefore, in agreement with Eq.~\eqref{eq:MassQuadrupoleD4}, only the combination $H_1+H_2$ enters in the definition of the mass quadrupole, and since the stress quadrupole is not independent of it, as shown in Eq.~\eqref{eq:StressQuadrupoleD4}, there is only one degree of freedom. Indeed, as previously discussed, one can find a coordinate transformation such that the entire metric depends on $H_1$ and $H_2$ only through the combination $H_1+H_2$. 

However, since the Hartle-Thorne metric describe the most generic rotating object in $D=4$, it contains the case in which such an object is a black hole. In this special scenario, no-hair theorems in $D=4$ state that the only vacuum solution with a regular horizon is the Kerr metric, which has a well defined tower of multipole moments only determined by the mass and spin~\cite{Geroch:1970cd,Hansen:1974zz}.
In the Hartle-Thorne formalism, the Kerr metric corresponds to the solution with $\zeta=1$ and, in terms of Eq.~\eqref{eq:QuadMetric}, this means that a Kerr black hole is reproduced by fixing
\begin{equation}\label{eq:Kerrcondition}
    H_1+H_2=1\ .
\end{equation}
From the amplitude perspective, this condition is satisfied by an infinite number of non-minimally coupled theories, since $H_1$ and $H_2$ can be chosen freely as long as the Kerr condition in Eq.~\eqref{eq:Kerrcondition} is satisfied. This can be associated to the degeneracy occurring in $D=4$, that reduces the number of physical degrees of freedom to one, still having two independent free parameters. 
In addition to non-minimal QFTs, the minimally-coupled theory, that gives rise to what we define the simplest metric\footnote{This name is inspired by the fact that a minimally-coupled theory is the simplest possible theory from a QFT perspective.}, also satisfies the Kerr condition. 
Thus, from the point of view of the effective QFT, the scattering amplitude computation unveils that the simplest vacuum solution in $D=4$ is the Kerr black hole. This is analogous to what was shown in \cite{Mougiakakos:2020laz, DOnofrio:2022cvn}, namely that there exists a reference frame in which the Schwarzschild-Tangherlini metric is induced by the minimally-coupled scalar action.

\subsection{Myers-Perry black holes in $D=5$}

Let us now consider higher dimensional solutions in General Relativity, and in particular Myers-Perry black holes in $D=5$~\cite{Myers:1986un}. First of all notice that in $D=5$ the number of Casimir associated to the rotation group $SO(4)$ is two, namely an object can rotate independently with respect to two independent axes. This means that the Myers-Perry metric in $D=5$ has two independent angular momenta, and since the classical metric is written in terms of $(x_1, y_1, x_2, y_2)$ coordinates, defined in such a way that the plane $(x_i, y_i)$ is orthogonal to the angular momentum $J_i$, we have to block diagonalize the spin tensor as
\begin{equation}
 S_{ij}=\begin{pNiceMatrix}[columns-width=auto]
        0 & J_1 & 0 & 0 \\
        -J_1 & 0 & 0 & 0 \\
        0 & 0 & 0 & J_2 \\
        0 & 0 & -J_2 & 0
    \end{pNiceMatrix}\ ,
\end{equation}
where the match between the physical angular momenta $J_i$ and the spin parameter of the Myers-Perry metric $a_i$ (written in Eq.~\eqref{eq:MPmetric}) is given by\footnote{Notice that with respect to the appendix~\ref{app:MP} $a_1=a$ and $a_2=b$.} 
\begin{equation}
   m\, a_i= \frac{D-2}{2} J_i\ .
\end{equation}
Hence, rewriting the metric in harmonic coordinates as shown in Appendix~\ref{app:MP}, and fixing the physical parameters as 
\begin{equation}\label{eq:MPcondition}
   H_1=\frac{3}{8}\qquad \text{and}\qquad H_2=\frac{15}{16}\ ,
\end{equation}
this metric matches exactly with our solution obtained from amplitudes up to the order of our expansion (3PM; we believe this is true at all PM orders).

Furthermore, we can see that in this case the degrees of freedom manifestly match, and $H_1$ and $H_2$ are independently fixed to a specific numerical value. From the discussion of the previous sections, this corresponds to having independent mass and stress multipole moments
\begin{equation}
    \mathbb{M}^{(2)}_{a_1a_2}\Big|_{d=4}^{\rm MP}=-\frac{9}{m^2}S_{a_1k}S_{a_2}{}^{k}\ , 
\end{equation}
\begin{equation}
    \mathbb{G}^{(2)}_{ij,a_1a_2}\Big|_{d=4}^{\rm MP}=-\frac{9}{2\, m^2}S_{(i|a_1}S_{|j)a_2}\ .
\end{equation}
This also shows that, from a QFT point of view, the Myers-Perry solution does not correspond to a minimally coupled theory. From this perspective, unlike the Kerr black hole, the Myers-Perry black hole is not the simplest vacuum solution in General Relativity for $D=5$, indeed Eq.~\eqref{eq:MPcondition} does not coincide with the minimal limit in~\eqref{eq:MinimalLimit}. Since the Myers-Perry solution is the natural generalization of the Kerr solution in arbitrary dimensions \cite{Emparan:2008eg}, one can conclude that, generally speaking, black holes are not the simplest vacuum solutions in arbitrary dimensions from a scattering amplitude perspective.

Finally, since when $J_1=J_2$ the full Myers-Perry metric features an enhanced symmetry becoming cohomogeneity-1~\cite{Myers:1986un}, one can observe that, in terms of multipole expansion, in the limit of equal angular momenta the mass quadruple vanishes~\cite{Heynen:2023sin}. Explicitly, it is possible to see that
\begin{equation}
   \lim_{J_1\rightarrow J_2} \mathbb{M}^{(2)}_{a_1a_2}\Big|_{d=4}^{\rm MP}N_{a_1a_2}=0\ , 
\end{equation}
while the stress quadrupole does not vanish. 
This is not only true for the Myers-Perry metric~\cite{Myers:1986un}, but actually for the general metric obtained from the amplitudes for any odd spacetime dimension $D$ when all the angular momenta are the same. This means that, in general, 
\begin{equation}
   \lim_{J_i\rightarrow J} \mathbb{M}^{(2)}_{a_1a_2}\Big|_{d=\text{even}}N_{a_1a_2}=0\ . 
\end{equation}
We interpret this as a generic property of the gravitational field sourced by a spinning point-like mass.

\subsection{Black rings}\label{sec:BlackRing}
As we already mentioned, in $D>4$ Myers-Perry black holes are not the only vacuum solutions with horizons. One solution beyond the Myers-Perry black holes is the black ring solution~\cite{Emparan:2006mm}, which has ring topology at the horizon (at variance with the Kerr and Myers-Perry solutions that instead have spherical topology).

Let us consider for simplicity the case with only one angular momentum, where $S_{21}=-S_{12}=J$ are the only non-vanishing components of the spin tensor. The original solution is written in terms of three parameters $(\mathcal{R}, \nu, \lambda)$ with $0<\nu\leq1$, which encode mass, spin, and shape of the ring. As shown in Appendix~\ref{app:BR}, we can replace $\mathcal{R}$ and $\nu$ in favor of $m$ and $J$, which are respectively the mass and the spin of the solution. In terms of $(m, J, \lambda)$, we can write the metric in harmonic coordinates in a PM expansion. 
For a generic value of $\lambda$, the black ring solution has a naked conical singularity unless
\begin{equation}
    \lambda=\frac{2\nu(m, J)}{1+\nu^2(m, J)}\ , \label{equilibrium}
\end{equation}
which in turn implies that mass and spin are not independent. 
This condition also corresponds to the existence of equilibrium solutions in the absence of external forces, and sets a lower bound on the angular momentum. 
However, we keep $0<\lambda\leq 1$ as a free parameter, thus describing a family of solutions, of which only one is free of naked conical singularities for a given $m$ and $J$. The Myers-Perry solution is then recovered from the black ring if $\lambda\rightarrow 1$, keeping fixed $m$ and $J$ (this is outside the equilibrium curve of the ring). 

With the aim of comparing the black ring metric with Eq.~\eqref{eq:QuadMetric}, consider the quadrupole structure described in~\eqref{eq:GenericMultipoleMetric}. Both mass and stress quadrupole moments have a dimension of a length squared (in $\hbar=c=1$ units), meaning that in $D=5$ they can be written in terms of dimensionful quantities $S^2/m^2$ or $Gm$\footnote{We recall that in higher dimension $[Gm]=L^{D-3}$.}. If a new fundamental length scale $\Lambda$ is present in the gravitational source, schematically the intrinsic quadrupole reads
\begin{equation}
    \mathbb{M}^{(2)}\sim \Lambda=  \sigma\, Gm\ ,
\end{equation}
where $\sigma$ is a dimensionless parameter.
This means that, if one writes $\Lambda$ in units of $Gm$, such non-spin-induced moment would enter at 2PM in the expansion of the metric, even though it is associated with a tree-level scattering amplitude. This is the case for the black ring solution, that has a ring topology and hence a non-vanishing intrinsic quadrupole moment even in the non-spinning limit.  
However, since the amplitude-based approach presented in this paper focuses on spin-induced multipole moments, we cannot match the metric at quadrupole level including intrinsic moments. We restrict in this case to 1PM order, in which only spin-induced moments are present and we expect to be able to fully reconstruct the metric and its spin-induced multipoles. 

Then, from the results of Appendix~\ref{app:BR}, fixing the physical parameters as 
\begin{equation}\label{eq:BRcondition}
        H_1=\frac{3}{4(1+\lambda)}\ ,\qquad H_2=\frac{3(6\lambda-1)}{8(1+\lambda)}\ ,
\end{equation}
we have verified the agreement of the black ring metric with the amplitude-based one, up to 1PM. Indeed, we can easily check that for $\lambda=1$ we recover exactly the expected coefficients $H_1$ and $H_2$ of the Myers-Perry solution in Eq.~\eqref{eq:MPcondition}.
As in the Myers-Perry black holes, in Eq.~\eqref{eq:BRcondition} for a given value of $\lambda$, the parameters $H_1$ and $H_2$ are independently fixed, and so also black rings in $D=5$ have independent mass and stress quadrupole moments. For completeness, their explicit expression is
\begin{align}
    \mathbb{M}^{(2)}_{a_1a_2}\Big|_{d=4}^{\text{BR}}&=-\frac{1}{m^2}\frac{18 \lambda}{1+\lambda} S_{a_1k}S_{a_2}{}^{k}+O(G m)\ , \\
    \mathbb{G}^{(2)}_{ij,a_1a_2}\Big|_{d=4}^{\text{BR}}&=-\frac{1}{m^2}\frac{9}{1+\lambda} S_{(i|a_1}S_{|j)a_2}+O(Gm)\ ,
\end{align}
where we are neglecting the intrinsic (\textit{i.e.} non-spin-induced) quadrupole moments. 
Given the extra parameter $\lambda$ (in addition to the mass and angular momentum), one could check whether there exists a particular black ring solution corresponding to the simplest metric generated by the minimal vertex of the QFT. However, its easy to verify that there is no value of $\lambda$ such that $H_1=1$ and $H_2=0$ at the same time. This shows that all black ring solutions with single angular momentum are generated by non-minimally coupled theories from the scattering amplitude perspective. 
%

\subsection{The simplest metric in any spacetime $D$ dimensions}
For completeness, we give here the explicit form of the simplest solution in generic spacetime $D$ dimensions, obtained from the minimal vertex in~\eqref{eq:MinimalVertex} and corresponding to $H_1=1$ and $H_2=0$:
\begin{equation}\label{eq:SimplestMetric}
    \begin{aligned}
        h_{00}^{(1, 2)}(r)&=\frac{2(D-3)}{D-2}\frac{r^2S_{k_1k_2}S^{k_1k_2}-(D-1)\, x^{k_1}x^{k_2}S_{k_1}{}^{k_3}S_{k_2k_3}}{mr^4}G \rho(r)\ ,\\
        h_{0i}^{(1, 2)}(r)&=0\ ,\\
        h_{ij}^{(1, 2)}(r)&=-\frac{2(D-3)}{(D-2)mr^4}\Bigg(-r^2(D-2)S_{ik}S_{j}{}^{k}+r^2S_{k_1k_2}S^{k_1k_2}\delta_{ij}\\
        &+(D-1)\, x^{k_1}x^{k_2}\Big((D-2)S_{ik_1}S_{jk_2}-S_{k_1}{}^{k_3}S_{k_2k_3}\delta_{ij}\Big)\Bigg)G\rho(r)\ . 
    \end{aligned}
\end{equation}
We stress again that the 1PM order contains all the physics that we want to capture, and we consider only the quadrupole term because monopole and dipole moments are uniquely fixed. In this regard the quadrupole moments of this solution are
\begin{align}
    \left.\mathbb{M}^{(2)}_{a_1a_2}\right|^{\rm simplest}&=-\frac{D-1}{m^2} S_{a_1k}S_{a_2}{}^{k}\ , \label{eq:MassQuadrupoleGenericDsimplest} \\
    \left.\mathbb{G}^{(2)}_{ij,a_1a_2}\right|^{\rm simplest}&=-\frac{(D-1)(D-2)}{m^2} S_{(i|a_1}S_{|j)a_2}\ .\label{eq:StressQuadrupoleGenericDsimplest}
\end{align}

As previously discussed, in $D=4$ this solution corresponds to the Kerr metric. However, when considering $D=5$, it does not match either the Myers-Perry metric or the black ring with a single angular momentum.
Since our analysis shows that the minimal solution is not a black hole, we leave it as an interesting open problem to identify whether an exact solution exists that corresponds to this metric, and possibly what is the matter content sourcing such solution, similarly to the Hartle-Thorne metric in $D=4$. Moreover, one could in principle still construct a black hole solution whose spin-induced part of each multipole matches the one of the simplest metric.
At tree-level the theory is linear and this could be done by superimposing various solutions and their corresponding moments. However, in general such superposition is invalid beyond the linear level, so constructing such solution  (if it exists) is highly nontrivial.

Another interesting possibility is the existence of different simplest solutions in $D>4$. As we briefly mentioned in Sec.~\ref{sec:MetricFromAmplitudes}, when we refer to spin-$s$ fields we are assuming a completely symmetric and trace-less field representation of the Lorentz group, but in higher dimensions there are more possible representations, and for example in $D=5$ we can build an action that couples an anti-symmetric field with gravity. The minimal vertex associated to this theory could be either equal to~\eqref{eq:MinimalVertex}, thus suggesting the fundamental nature of the simplest metric, or different, leading to an another simplest solution that can be compared to Myers-Perry or other black hole solutions. Likewise, one could consider a theory for a minimally coupled field with spin $s>1$ and check whether the simplest metric is universal, at least up to the quadrupole order.

Finally, it is important to stress that Eq.~\eqref{eq:FinalQuadVertex} (and therefore the most generic metric in $D$ dimensions) could also be derived without the use of scattering amplitudes, but simply by constructing the most generic stress-energy tensor compatible to its symmetries in momentum space. However, having an underlying QFT allows us to identify the simplest solution among the general family, and gives to us the possibility to interpret black holes from a different point of view. 
The fact that in $D=4$ the simplest solution corresponds to the Kerr metric seems to resemble the no-hair theorem in a QFT language, in which the absence of hair corresponds to the absence of extra coupling in the effective action. 
In $D>4$ the story is more complicated, and the absence of black hole uniqueness theorems seems to spoil the connection between simplest metrics and black holes solutions, although our work suggests further investigations along this line.

\section{Conclusions}\label{sec:conclusions}
In this work we computed the vacuum solution describing the metric of a generic rotating object in arbitrary dimensions, up to spin-induced quadrupole order and within a PM expansion. Our computation is based on the
the classical contributions of scattering amplitudes describing the emission of gravitons out of massive spin-1 particles and provides valuable insights into the nature of spinning compact objects in arbitrary dimensions.

In the context of $D=4$ spacetime dimensions, our computations have successfully recovered the PM expansion of the well-known vacuum Hartle-Thorne solution, describing the metric of a spinning compact object in General Relativity up to second order in the angular momentum. The Kerr metric describing a spinning black hole is a special case arising for a specific choice of the mass quadrupole moment. At the level of the effective quantum field theory, this choice corresponds to the case of a Proca field minimally coupled to gravity. This reaffirms black holes as the simplest solutions to General Relativity in four dimensions also from a scattering-amplitude perspective.

However, our investigation reveals that the situation is different and richer in higher ($D>4$) dimensions. Here, through our scattering amplitude analysis up to 2-loop calculations, we have obtained the generic solution including quadrupole-moments terms quadratic in the object's angular momentum.
Interestingly, at variance with the $D=4$ case, we have found that the solution depends on \emph{two} quadrupole moment parameters, namely the standard mass quadrupole and a new stress quadrupole; the latter does not exist in $D=4$ and was previously missed in the analysis of~\cite{Heynen:2023sin}.
We have successfully identified and characterized different $D>4$ black-hole solutions, such as the Myers-Perry and the black ring metrics, as particular cases obtained for specific choices of the parameters of our general PM solution.
Notably, unlike the four-dimensional case, none of these solutions corresponds to the choice of minimal couplings in the effective action, highlighting that the nature of black holes in higher dimensions is more intricate. In this direction our work could be important to extend the analysis of 4 and 5-point classical gravitational scattering amplitudes involving spinning bodies~\cite{Siemonsen:2019dsu, Vines:2017hyw, Vines:2018gqi, Guevara:2018wpp, Luna:2023uwd,Brandhuber:2023hhl, DeAngelis:2023lvf} to higher dimensions.

It would be interesting to assess whether this property is somehow linked to the absence of a black-hole uniqueness and no-hair theorem in higher-dimensional General Relativity. Nevertheless, it is known that the Myers-Perry solution is the unique stationary, non-extremal, asymptotically flat, vacuum black hole solution with spherical topology \cite{Emparan:2008eg, Morisawa:2004tc}, which are exactly the hypotheses behind our simplest metric. This means that if the Myers-Perry solution is not the simplest one, strictly speaking no other solution can be within our framework. 
However, an interesting follow-up is to understand whether the concept of simplest metric is universal in $D>4$, for example by constructing the metric arising from a minimally coupled theory other than Proca, in this case it could happen that another simplest solution matches the Myers-Perry one.

Even though we focused on the case of solutions with spin-induced multipole moments, it should be possible to extend our framework to describe generically deformed compact objects 
(an example being the black ring solution that has intrinsic moments in addition to the spin-induced ones), also featuring higher-order multipole moments or moments that break the Kerr symmetries (e.g., current quadrupoles, mass and stress octupoles that break the equatorial symmetry, or generically moment tensors that break the axisymmetry)~\cite{Raposo:2018xkf,Bena:2020see,Bianchi:2020bxa,Bena:2020uup,Bianchi:2020miz,Fransen:2022jtw}.

Another interesting avenue of exploration would be to identify the solution corresponding to the minimal coupling in $D>4$ at the full nonlinear level. Our results suggest that such a solution (if it exists) should be stellar-like, \textit{i.e.} it requires matter fields coupled to gravity, or perhaps 
it corresponds to black holes with complex topology as those obtained by ``superimposing'' different black hole solutions using the inverse scattering method, as in the case of the black saturn~\cite{Elvang:2007rd}, multi-ring, or other~\cite{Emparan:2010sx,Elvang:2007hg}, possibly yet unknown, solutions~\cite{Emparan:2008eg}.
The study of compact objects other than black holes and their multipolar structure in higher dimensions is seldom explored and worth investigating to address this question.

We conclude by highlighting that the discovery of stress multipoles opens several interesting avenues for future investigation, both at the phenomenological level and for a more fundamental understanding of gravity. Although we have shown that the stress moments do not exist in $D=4$ General Relativity, it would be very interesting to assess whether this happens also for other gravity theories. We expect that other gravitational theories with more degrees of freedom (e.g. massive gravity or generic metric theories propagating up to six polarizations at the linearized level) should indeed have non-vanishing stress multipoles. If this is the case, it would be interesting to understand the phenomenology associated to these new moments. This might also change the way in which we build stationary solutions beyond General Relativity or perform tests of exotic compact objects (see, e.g,~\cite{Loutrel:2022ant}). Furthermore, in higher dimensions these considerations have to be taken into account already within General Relativity, potentially impacting on the way in which we look for new solutions and study their linearized dynamics. 
For example, even for a spherical object the new multipole moments can be induced by an external tidal field, giving rise to \emph{stress Love numbers}, associated with tensor perturbations of compact objects in $D>4$~\cite{Kodama:2003jz}, which are indeed absent in $D=4$ (see~\cite{Kol:2011vg,Cardoso:2019vof,Hui:2020xxx} for a discussion of the tidal Love numbers in higher dimensions). 

Finally, even though the stress multipoles could have been found within a classical General Relativity framework, considering scattering amplitudes was really key to identify their existence. In particular, we realized that working in momentum space it is relatively easy to write down the most generic stress-energy tensor at arbitrary high-order in the multipole expansion, at variance with working in position space (the natural setting of General Relativity), in which deriving the most generic expression of $T_{\mu\nu}$ would be more and more tedious as the order of multipoles increases. This is a perfect example of how recovering classical gravity from scattering amplitudes not only can give us insights on the phenomenology, but can also open new perspectives to deepen our understanding of gravity at a more fundamental level.

\textbf{Note added in v2:} After the completion of this work, we became aware of~\cite{Amalberti:2023ohj}, which performs a multi- pole expansion of the long-wavelength effective action for radiative sources in higher dimensions. The authors find a new set of Weyl-type moments, which coincide with the stress moments of a stationary object defined in this work.

\begin{acknowledgments}
We thank George Pappas for interesting correspondence, and 
Roberto Emparan, Leonardo Gualtieri, Jef Heynen and Daniel Mayerson for useful comments on the draft.
This work is partially supported by the MIUR PRIN Grant 2020KR4KN2 “String
Theory as a bridge between Gauge Theories and Quantum Gravity”, by the FARE programme (GW-NEXT, CUP:~B84I20000100001), by the EU Horizon 2020 Research and Innovation Programme under the Marie Sklodowska-Curie Grant Agreement No. 101007855, and by the INFN TEONGRAV initiative.
\end{acknowledgments}

\appendix

\section{Quadrupole moments in four and five spacetime dimensions}\label{app:multipole}

In this appendix we perform in detail the decomposition in irreducible $SO(d)$  representations of the constant tensor $\mathcal{G}_{ij, ab}$ which enters the expansion in Eq.~\eqref{eq:gammaMetric} and it is relevant for the analysis of the quadrupole moments. As observed in subsection~\ref{sec:multipolemoments}, the tensor is symmetric in the indices $ij$ and $ab$, and it is also traceless in $ab$. We will consider explicitly the five-dimensional case, but the same results apply in any dimension $D>4$. To highlight the difference with the four-dimensional case, we will first quickly review how the latter works, and then move to $D=5$.

In four dimensions the tensor $\mathcal{G}_{ij, ab}$ belongs to the $({\bf 5} \oplus {\bf 1})\otimes {\bf 5}$ of $SO(3)$. This decomposes in irreducible representations as 
\begin{equation}
    ({\bf 5} \oplus {\bf 1})\otimes {\bf 5} = {\bf 9} \oplus {\bf 7} \oplus 2 \times {\bf 5} \oplus {\bf 3} \oplus {\bf 1}
\end{equation}
which correspond to symmetric traceless tensors with 4, 3, 2, 1 and 0 indices of $SO(3)$, respectively. It is straightforward to recognize these representations as the ones of the tensors $\mathcal{G}^{(4)}$, $\mathcal{G}^{(6)}$, $\mathcal{G}^{(1)}$, $\mathcal{G}^{(3)}$, $\mathcal{G}^{(5)}$ and $\mathcal{G}^{(2)}$, respectively, that appear in Eq.~\eqref{eq:gammaMetricquadrupole}. 

We can now move to five dimensions. In this case the indices belong to $SO(4)$, which is isomorphic to $SU(2) \times SU(2)$. We can then label the representations 
in terms of the ones of each of the two $SU(2)$'s, and in particular the tensor $\mathcal{G}_{ij, ab}$  belongs to the $(({\bf 3}, {\bf 3})\oplus ({\bf 1},{\bf 1}))\otimes ({\bf 3},{\bf 3})$, which decomposes in irreducible representations as
\begin{equation}
\begin{aligned}
& (({\bf 3}, {\bf 3})\oplus  ({\bf 1},{\bf 1}))\otimes ({\bf 3},{\bf 3})=  ({\bf 5}, {\bf 5})\oplus ({\bf 5}, {\bf 3}) \oplus ({\bf 3}, {\bf 5})\\ & \oplus  ({\bf 5}, {\bf 1})\oplus ({\bf 1}, {\bf 5})\oplus 2\times ({\bf 3}, {\bf 3})\oplus  ({\bf 3}, {\bf 1})\oplus   ({\bf 1}, {\bf 3})\oplus ({\bf 1}, {\bf 1})\ . 
\end{aligned}
\end{equation}

We want to identify such representations with the index structure and constraints of the tensors in Eq.~\eqref{eq:gammaMetricquadrupoleD5}. The $({\bf 5}, {\bf 5})$ representation corresponds to the four-index symmetric traceless tensor $\mathcal{G}^{(4)}$, while the $({\bf 3}, {\bf 3})$'s correspond to the two-index symmetric traceless tensors $\mathcal{G}^{(1)}$ and $\mathcal{G}^{(3)}$, and the $({\bf 1}, {\bf 1})$ is the singlet $\mathcal{G}^{(2)}$. The $({\bf 3}, {\bf 1})\oplus   ({\bf 1}, {\bf 3})$ identifies the antisymmetric tensor $\mathcal{G}^{(5)}$. The remaining representations, namely  $({\bf 5}, {\bf 3}) \oplus ({\bf 3}, {\bf 5})$ and $({\bf 5}, {\bf 1})\oplus ({\bf 1}, {\bf 5})$, correspond to tensors with mixed symmetry. 

The tensor $\mathcal{G}^{(6)}$ in the $({\bf 5}, {\bf 3}) \oplus ({\bf 3}, {\bf 5})$ is the one identified in~\cite{Heynen:2023sin} as the ASTF tensor $\mathcal{H}_{ab, cd}$ with four indices, antisymmetric in the first two and symmetric traceless in the last two. Such tensor, being irreducible, satisfies the additional constraints\footnote{To avoid overburderning with notation, in this appendix we leave the STF, ASTF, and RSTF notation implicit, for instance $\mathcal{G}^{(6)}_{ab, cd}\equiv \mathcal{G}^{(6)}_{\{ab, cd\}_{\rm ASTF}}$ and $\mathcal{G}^{(7)}_{ab, cd}\equiv \mathcal{G}^{(7)}_{\{ab, cd\}_{\rm RSTF}}$.}
\begin{equation} \label{eq:grouptheoryconstraints}
    \delta^{bc} \mathcal{G}^{(6)}_{ab, cd} = 0 \qquad \epsilon^{abce} \mathcal{G}^{(6)}_{ab,cd} =0 \ .  
\end{equation}
To show that these constraints lead to 30 surviving components\footnote{Note that~\cite{Heynen:2023sin} ascribes 40 components to this tensor. We believe this is a mistake. Indeed, the 40 components corresponds to the 30 components of $\mathcal{G}^{(6)}$ plus the 10 components of the tensor $\mathcal{G}^{(7)}$ described below.} we can start by observing that the number of components of the tensor before imposing the constraints is $6 \times 9 = 54$. The first constraint in Eq.~\eqref{eq:grouptheoryconstraints} removes 15 components, because both the indices $a$ and $d$ can take any value but the trace in $ad$ of the constraint is identically zero. The second constraint in Eq.~\eqref{eq:grouptheoryconstraints} removes 9 additional components. Indeed, given the previous constraint, $\epsilon^{abce} \mathcal{G}^{(6)}_{ab,cd}$ is symmetric traceless in $ed$. The fact that it is traceless is due to the symmetry in $cd$, while the fact that it is symmetric can be seen using the Fierz identity
\begin{equation}
    \epsilon^{[abce} \mathcal{G}^{(6)}_{ab,c}{}^{d]} = 0  \ ,
\end{equation}
which implies that the antisymmetric part of $\epsilon^{abce} \mathcal{G}^{(6)}_{ab,cd}$ vanishes identically.
We are thus left with $54 - 15 - 9 =30$ components.

The tensor $\mathcal{G}^{(7)}$ in the $({\bf 5}, {\bf 1}) \oplus ({\bf 1}, {\bf 5})$, which is missing in~\cite{Heynen:2023sin}, is antisymmetric in both the pairs $ab$ and $cd$, which makes in total 36 components. Besides, it satisfies the constraints 
\begin{equation} \label{eq:grouptheoryconstraints2}
    \delta^{bc} \mathcal{G}^{(7)}_{ab, cd} = 0 \qquad \epsilon^{abce} \mathcal{G}^{(7)}_{ab,cd} =0 \ .
\end{equation}
Using arguments similar to the ones above, one can show that 
the first constraint removes 16 components and the second removes 10 components, leaving in total $36 -16 - 10 = 10$ components.

The analysis performed here in five dimensions can be generalized to any $D>4$, implying that only in four dimensions one can choose a gauge such that $\gamma_{ij}=0$ as in Eq.~\eqref{eq:gammaijvanishes}.

\section{Hartle-Thorne metric in harmonic coordinates}\label{app:Hartle-Thorne}

The most generic rotating object in $D=4$ is described up to spin-induced quadrupole order by the Hartle-Thorne metric, which is exact in $G$~\cite{Hartle:1967he, Hartle:1968si}. Given a spherical set of coordinates $(t, r, \theta, \phi)$, the explicit expression of the metric up to 3PM order is
\begin{equation}\label{eq:HartleThorneMetric}
    \begin{aligned}
    g_{tt}&=1-\frac{2 G m}{r}+\frac{a^2 G m \zeta (3
   \cos (2 \theta )+1)}{2 r^3}\\
   &+\frac{a^2G^2 m^2 ((3
   \zeta -2) \cos (2 \theta )+\zeta -2)}{2 r^4} +\frac{a^2G^3m^3
   (4 \zeta -11)  (3 \cos (2 \theta )+1)}{7 r^5}+O(G^4, a^3)\ ,\\
    g_{t\phi}&=+ \frac{2 a G m \sin^2(\theta)}{r}+O(G^4, a^3)\ ,\\
    g_{rr}&=-1- \frac{2 Gm}{r}+\frac{(a^2Gm \zeta  (3 \cos (2 \theta )+1)) }{2
   r^3}- \frac{4 G^2m^2}{r^2}-\frac{a^2G^2m^2
   (-5 \zeta -3 (5 \zeta -8) \cos (2 \theta )+4) }{2
   r^4}\\
   &- \frac{8 G^3m^3}{r^3}-\frac{a^2G^3m^3
   (-60 \zeta -9 (20 \zeta -27) \cos (2 \theta )+25)}{7
   r^5}+O(G^4, a^3)\ ,\\
   g_{\theta\theta}&=-r^2+\frac{a^2G m\zeta   (3 \cos (2 \theta
   )+1)}{2 r}+\frac{a^2G^2m^2 (5 \zeta -1) (3
   \cos (2 \theta )+1)}{4 r^2}\\
   &+\frac{18 a^2G^3m^3 (\zeta -1)
    (3 \cos (2 \theta )+1)}{7 r^3}+O(G^4, a^3)\ ,\\
    g_{\phi\phi}&=-r^2 \sin ^2(\theta ) +\frac{a^2Gm \zeta  (3 \cos (2 \theta
   )+1) \sin ^2(\theta )}{2 r}+\frac{\left(a^2G^2m^2 (5 \zeta -1) (3 \cos (2 \theta )+1) \sin
   ^2(\theta )\right) a^2}{4 r^2}\\
   &+\frac{18 \left(a^2G^3m^3 (\zeta -1) (3 \cos (2 \theta )+1) \sin
   ^2(\theta )\right)}{7r^3}+O(G^4, a^3)\ ,
    \end{aligned}
\end{equation}
where $a=J/m$ is the angular momentum per unit mass and $\zeta$ is the quadrupole mass moment normalized such that for $\zeta=1$ we have the Kerr limit. 

In order to compare such metric with the one derived by scattering amplitudes, we have to impose the harmonic gauge. The reason why this reference frame is particularly suitable is twofold: on one hand this is a ``universal gauge'', in the sense that there exists a gauge fixing term at the level of the action such that every metric can be put in this frame, and on the other hand it is classically defined by the covariant d'Alambertian of the associated Cartesian coordinates, that makes it possible to define the transformation on both the original and the transformed metric. Defining the spherical harmonic coordinates as $(T, R, \Theta, \Phi)$, and the associated Cartesian coordinates as
\begin{equation}
    \begin{cases}
    x=R\sin(\Theta)\cos(\Phi)\ ,\\
    y=R\sin(\Theta)\sin(\Phi)\ ,\\
    z=R\cos(\Theta)\ ,
    \end{cases}
\end{equation}
the equation that the metric in harmonic spherical coordinates has to satisfy is 
\begin{equation}\label{eq:HarmonicBox}
    g^{\mu\nu}D_\mu \partial_\nu (T, x, y, z)=0\ ,
\end{equation}
where each coordinate is treated as a scalar. 
Noticing that the metric in~\eqref{eq:HartleThorneMetric} does not depend on the azimuthal angle $\phi$, we can define a coordinate transformation as 
\begin{equation}\label{eq:HarmonicCoordTransfD4}
  T=t\ ,\qquad  R=r(R, \Theta)\ , \qquad \Theta=\theta(R, \Theta)\ ,\qquad \Phi=\phi \ . 
\end{equation}

We can now define an ansatz for the relation between the two set of coordinates as
\begin{equation}
    r(R, \Theta)=R\sum_{i=0}^{n\text{PM}}\left(\frac{G m}{R}\right)^i\sum_{j=0}^{\lfloor n\text{Pole}/2\rfloor}\left(\frac{a}{R}\right)^{2j}\sum_{k=0}^{j}\mathcal{C}_{i, 2j, k}^{(R)}P_{2k}(\cos\Theta)\ ,
\end{equation}
\begin{equation}\label{eq:ThetaD4HarmonicTransf}
    \cos\theta(R, \Theta)=\cos(\Theta)\Bigg(1+\sum_{i=0}^{n\text{PM}}\left(\frac{G m}{R}\right)^i\sum_{j=1}^{\lfloor n\text{Pole}/2\rfloor}\left(\frac{a}{R}\right)^{2j}\sum_{k=0}^{j}\mathcal{C}_{i, 2j, k}^{(\Theta)}P_{2k}(\cos\Theta)\Bigg)\ ,
\end{equation}
where $P_n$ are the Legendre polynomials and $\lfloor\cdot\rfloor$ stands for the integer part.
The ansatz is motivated as follows:
\begin{itemize}
    \item In the limit in which $R\rightarrow+\infty$ the two reference frames must coincide, which implies $\mathcal{C}^{(R)}_{0, 0, 0}=1$.
    \item In the non-spinning limit ($a\rightarrow 0$), the Schwarzschild metric must be recovered, and so in~\eqref{eq:ThetaD4HarmonicTransf} for $a=0$ we have to impose $\theta=\Theta$.
    \item In the original metric at a given spin power, the order of the Legendre polynomials is always equal or lower than the spin power itself, hence we set $k<j$.
    \item Since we have to respect the time-reversal symmetry in which $t\rightarrow -t$ and $\Phi\rightarrow-\Phi$, only even powers of the angular momentum are allowed.
\end{itemize}

Finally, performing the coordinate transformation and imposing equation~\eqref{eq:HarmonicBox}, we end up with 
\begin{equation}\label{eq:MapHTcoeffs}
\begin{gathered}
    \mathcal{C}^{(R)}_{0, 0, 0}=1\ , \quad \mathcal{C}^{(R)}_{0, 2,2}=0\ ,\quad \mathcal{C}^{(R)}_{0, 2,0}=0\ ,\quad \mathcal{C}^{(\Theta)}_{0, 2,2}=0\ ,\quad \mathcal{C}^{(\Theta)}_{0, 2,0}=0\ ,\\
    \mathcal{C}^{(R)}_{1, 0, 0}=1\ , \quad \mathcal{C}^{(\Theta)}_{1, 2, 2}=-\mathcal{C}^{(R)}_{1, 2, 2}\ , \quad \mathcal{C}^{(\Theta)}_{1, 2, 0}=\mathcal{C}^{(R)}_{1, 2, 0}\ ,\\
    \mathcal{C}^{(R)}_{2, 0, 0}=0\ ,\quad \mathcal{C}^{(\Theta)}_{2, 2, 2}=\frac{\zeta-4}{3}\ ,\quad \mathcal{C}^{\Theta}_{2, 2, 0}=\frac{4-\zeta}{3}\ ,\quad \mathcal{C}^{(R)}_{2, 2, 2}=\frac{7}{3}\ ,\quad \mathcal{C}^{(R)}_{2, 2, 0}=-\frac{1}{3}\ ,\\
    \mathcal{C}^{(\Theta)}_{3, 2, 2}=\frac{64}{9}-\mathcal{C}^{(R)}_{1, 2, 2}+\frac{2}{3}\mathcal{C}^{(R)}_{3, 2, 2}\ , \quad \mathcal{C}^{(\Theta)}_{3, 2, 0}=-\frac{64}{9}+\mathcal{C}^{(R)}_{1, 2, 2}-\frac{2}{3}\mathcal{C}^{(R)}_{3, 2, 2}\ , \quad \mathcal{C}_{3, 2, 0}^{(R)}=\frac{2}{3}+\frac{3}{5}\mathcal{C}^{(R)}_{1, 2, 0}\ .
\end{gathered}
\end{equation}
Once the coordinate transformation is fixed, we can find the metric in harmonic gauge at the chosen perturbative order by means of~\eqref{eq:HarmonicCoordTransfD4}.
We notice that the coordinate transformation has some gauge redundancies since there are coefficients which are unfixed. At 1PM and 2PM the redundancy is parameterized by two coefficients, namely $\mathcal{C}^{(R)}_{1, 2, 0}$ and $\mathcal{C}^{(R)}_{1, 2, 2}$. At 3PM a new redundancy arises by means of the unfixed coefficients $\mathcal{C}^{(R)}_{3, 0, 0}$ and $\mathcal{C}^{(R)}_{3, 2, 2}$. While the full 3PM result is given in the attached Mathematica notebook \cite{AnchillaryFiles}, at 1PM the explicit expression of the Hartle-Thorne metric in harmonic gauge reads
\begin{equation}\label{eq:HartleThorneMetricHarmonic}
    \begin{aligned}
    g_{tt}&=1-\frac{2 G m}{R}+\frac{a^2 G m \zeta (3
   \cos (2 \Theta )+1)}{2 R^3}+O(G^2, a^3)\ ,\\
    g_{t\Phi}&=\frac{2 a G m \sin^2(\Theta)}{R}+O(G^2, a^3)\ ,\\
    g_{RR}&=-1- \frac{2 Gm}{R}+a^2Gm\frac{   8\mathcal{C}^{(R)}_{1, 2, 0}+(3 \cos (2 \Theta )+1)(\zeta+2\mathcal{C}^{(R)}_{1, 2, 2}) }{2
   R^3}+O(G^2, a^3)\ ,\\
   g_{\Theta\Theta}&=-R^2-2GmR+a^2G m\frac{\zeta (3\cos(2\Theta)+1)+\mathcal{C}^{(R)}_{1, 2, 2}(3\cos(2\Theta)-1)-4\mathcal{C}^{(R)}_{1, 2, 0}}{2 R}+O(G^2, a^3)\ ,\\
    g_{R\Theta}&=-\frac{3Gma^2\sin^2(\Theta)\mathcal{C}^{(R)}_{1, 2, 2}}{4R^2}+O(G^2, a^3)\ ,\\
    g_{\Phi\Phi}&=-R^2\sin^2(\Theta)-2GmR\sin^2(\Theta) \\
    &+a^2Gm\sin^2(\Theta)\frac{\zeta (3\cos(2\Theta)+1)-4\mathcal{C}^{(R)}_{1, 2, 0}+2\mathcal{C}^{(R)}_{1, 2, 2}}{2 R}+O(G^2, a^3)\ ,\\
    \end{aligned}
\end{equation}
and moving to Cartesian coordinates such metric can be directly compared with the amplitude-based one. 

This perturbative approach to the harmonic coordinate transformation can be compared directly with~\cite{Aguirregabiria:2001vk}, in which a similar procedure is described. Considering only terms up to 3PM and spin-square orders, we find disagreement with the number of gauge redundancies present in the metric after the transformation. In particular at 1PM we find two free parameters, while in the mentioned work there is only one present. Our result is also confirmed by the amplitude approach, in which as we showed in subsection~\ref{sec:HartleThorneComparison} there are two gauge degrees of freedom coming from the dressed vertex in $D=4$.

\section{Myers-Perry black holes in $D=5$ in harmonic coordinates}\label{app:MP}

Myers-Perry black holes are a class of stationary vacuum solutions of General Relativity built to be the generalization of the Kerr metric in arbitrary dimensions~\cite{Myers:1986un}. In $D=5$ the metric reads
\begin{equation}\label{eq:MPmetric}
\begin{aligned}
ds^2=dt^2&-\frac{\mu}{\Sigma}\left(dt+a\, \sin^2\theta\, d\phi_1+b\, \cos^2\theta \, d\phi_2\right)^2-\frac{r^2\Sigma}{\Pi-\mu r^2}dr^2\\
&-\Sigma d\theta^2-(r^2+a^2)\sin^2\theta\, d\phi_1^2-(r^2+b^2)\cos^2\theta\, d\phi_2^2\ ,
\end{aligned}
\end{equation}
where 
\begin{equation}
    \Sigma=r^2+a^2\cos^2\theta+b^2\sin^2\theta\ ,\qquad \Pi=(r^2+a^2)(r^2+b^2)\ , 
\end{equation}
and $a$ and $b$ are two independent angular momenta and 
\begin{equation}
    \mu=\frac{16 \pi G m}{(D-2)\Omega_{D-2}}
\end{equation}
with $\Omega_D$ the surface of a $D$-sphere. 
As we did in the $D=4$ case, we can define a set of Cartesian harmonic coordinates as
\begin{equation}\label{eq:HarmCartesianD5}
    \begin{cases}
    x_1=R\sin(\Theta)\cos(\Phi_1)\ ,\\
    y_1=R\sin(\Theta)\sin(\Phi_1)\ ,\\
    x_2=R\cos(\Theta)\cos(\Phi_1)\ ,\\
    y_2=R\cos(\Theta)\sin(\Phi_1)\ ,
    \end{cases}
\end{equation}
related to the original coordinates through
\begin{equation}\label{eq:HarmonicCoordinatesD5}
  T=t\ ,\qquad  R=r(R, \Theta)\ , \qquad \Theta=\theta(R, \Theta)\ ,\qquad \Phi_1=\phi_1\ ,\qquad \Phi_2=\phi_2 \ , 
\end{equation}
and such that 
\begin{equation}\label{eq:BoxD5}
    g^{\mu\nu}D_\mu \partial_\nu  (T, x_1, y_1, x_2, y_2)=0\ .
\end{equation}

Motivated by the same reasons of appendix~\ref{app:Hartle-Thorne}, the ansatz for the perturbative coordinate transformation is the following
\begin{equation}\label{eq:MPcoordinatetransf}
\begin{aligned} 
    &r(R, \Theta)=R\sum_{i=0}^{n\text{PM}}\left(G m\rho(R)\right)^i\sum_{j=0}^{\lfloor n\text{Pole}/2\rfloor}\sum_{k=0}^{j}\mathcal{C}_{i, 2j, 2k}^{(R)}\Bigg(\left(\frac{a}{R}\right)^{2j}P_{2k}(\cos\Theta)+\left(\frac{b}{R}\right)^{2j}P_{2k}(\sin\Theta)\Bigg)\\
    &+R\sum_{i=2}^{nPM}\left(G m\rho(R)\right)^i\log(mG\rho(R))\sum_{j=0}^{\lfloor n\text{Pole}/2\rfloor}\sum_{k=0}^{j}\mathcal{C}_{i, 2j, 2k}^{(\text{log},R)}\Bigg(\left(\frac{a}{R}\right)^{2j}P_{2k}(\cos\Theta)+\left(\frac{b}{R}\right)^{2j}P_{2k}(\sin\Theta)\Bigg)\ , \\
    &\cos\theta(R, \Theta)=\cos\theta\sum_{i=0}^{n\text{PM}}\left(G m\rho(R)\right)^i\sum_{j=0}^{\lfloor n\text{Pole}/2\rfloor}\sum_{k=0}^{j}\Bigg(\mathcal{C}_{i, 2j, 2k}^{(\Theta, a)}\left(\frac{a}{R}\right)^{2j}P_{2k}(\cos\Theta)+\mathcal{C}_{i, 2j, 2k}^{(\Theta, b)}\left(\frac{b}{R}\right)^{2j}P_{2k}(\sin\Theta)\Bigg)\\
    &+R\sum_{i=2}^{nPM}\left(G m\rho(R)\right)^i\log(mG\rho(R))\!\!\!\!\sum_{j=0}^{\lfloor n\text{Pole}/2\rfloor}\!\!\!\sum_{k=0}^{j}\Bigg(\mathcal{C}_{i, 2j, 2k}^{(\text{log},\Theta, a)}\left(\frac{a}{R}\right)^{2j}P_{2k}(\cos\Theta)+\mathcal{C}_{i, 2j, 2k}^{(\text{log},\Theta, b)}\left(\frac{b}{R}\right)^{2j}P_{2k}(\sin\Theta)\Bigg)\ .
    \end{aligned}
\end{equation}
Notice that differently from the $D=4$ case, here we have to consider logarithmic pieces starting from 2PM. This is consistent with the amplitude based approach in which we observe the presence of singularities in $D=5$ even in the harmonic gauge, renormalized by the insertion of higher-loop vertices giving rise to logarithmic terms. Moreover the coordinate transformation in Eqs.~\eqref{eq:MPcoordinatetransf} takes into account the symmetries of the original metric, such that the symmetry $a\rightarrow b$ and $\Theta\rightarrow \Theta+\pi/2$.

As we did in the Hartle-Thorne case, replacing Eq.~\eqref{eq:MPcoordinatetransf} in the harmonic equation, one is able to fix the coefficients of the coordinate transformation. In this case the equivalent of Eq.~\eqref{eq:MapHTcoeffs} for the Myers-Perry solution is way more involved, and since there is no physical information in it, we provide it in the attached Mathematica notebook \cite{AnchillaryFiles}. However, it is worth mentioning that coherently to what we know from the amplitude calculation, the coordinate transformation develops two independent gauge redundancies at orders $O(Ga^2)$, and two more at orders $O(G^2a^0)$ and $O(G^2a^2)$. In particular the redundancies at 2PM are related to the logarithmic terms and are the classical counterpart of the higher-loop vertex in Eq.~\eqref{eq:CounterTermVertex}.

Finally, we report the 3PM Myers-Perry metric in harmonic coordinates in the attached Mathematica file \cite{AnchillaryFiles}, giving here for the sake of clarity only the expression at 1PM in the case in which $\mathcal{C}_{1, 2, 0}^{(R)}=0$ and $\mathcal{C}_{1, 2, 2}^{(R)}=-2/9$ such that $g_{R\Theta}=0$, resulting in the simplified expression
\begin{equation}\label{eq:MPmetricHarmonic}
    \begin{aligned}
    g_{tt}&=1-\frac{8 G m}{3\pi R^2}+\frac{8 G m(a^2-b^2) \cos (2 \Theta )}{3\pi R^4}+O(G^2, a^3)\ ,\\
    g_{t\Phi_1}&=-\frac{8 a G m \sin^2(\Theta)}{3\pi R^2}+O(G^2, a^3)\ ,\\
    g_{t\Phi_2}&=-\frac{8 b G m \cos^2(\Theta)}{3\pi R^2}+O(G^2, a^3)\ ,\\
    g_{RR}&=-1-\frac{4Gm}{3\pi R^2}+\frac{4Gm(a^2-b^2)\cos(2\Theta)}{3\pi R^4}+O(G^2, a^3)\ ,\\
    g_{\Theta\Theta}&=-\frac{4Gm}{3\pi}-R^2+\frac{4Gm}{9\pi R^2}\Big(a^2+b^2+3(a^2-b^2)\cos(2\Theta)\Big)+O(G^2, a^3)\ ,\\
    g_{\Phi_1\Phi_1}&=-\frac{4Gm\sin^2(\Theta)}{3\pi}-R^2\sin^2(\Theta)-\frac{2Gm\sin^2(\Theta)}{9\pi R^2}\Big(a^2+b^2-3(3a^2-b^2)\cos(2\Theta)\Big)+O(G^2, a^3)\ ,\\
    g_{\Phi_2\Phi_2}&=-\frac{4Gm\cos^2(\Theta)}{3\pi}-R^2\cos^2(\Theta)-\frac{2Gm\cos^2(\Theta)}{9\pi R^2}\Big(a^2+b^2-3(a^2-3b^2)\cos(2\Theta)\Big)+O(G^2, a^3)\ ,\\
    g_{\Phi_1\Phi_2}&=-\frac{2abGm\sin^2(2\Theta)}{3\pi R^2}+O(G^2, a^3)\ .\\
    \end{aligned}
\end{equation}
To the best of our knowledge, there is no other derivation of the Myers-Perry metric in such coordinates in the literature to be compared with Eq.~\eqref{eq:MPmetricHarmonic}. However, moving to Cartesian coordinates, we can directly match it with the amplitude-based metric, finding a perfect agreement up to the order of our calculation, \textit{i.e.} 3PM. Even though all the physical information is contained in the tree-level part of the metric, higher-order calculations are an important and non-trivial consistency check that this approach passes perfectly.

\section{Black rings with one angular momentum in harmonic coordinates}\label{app:BR}

It is known that in $D>4$ there is no black hole uniqueness theorem, meaning that if one considers non-spherical topologies, the Myers-Perry black hole is not the only solution with a horizon. Black rings are indeed another black hole solution in $D=5$~\cite{Emparan:2001wn}. Limiting to the case of one angular momentum along the $\varphi_1$-direction, in the so-called black ring coordinates $(x, y, \varphi_1, \varphi_2)$ the explicit metric reads \cite{Emparan:2001wn}
\begin{equation}\label{eq:BRoriginalMetric}
\begin{aligned}
ds^2=&-\frac{A(y)}{A(x)}\left(dt-C\, \mathcal{R}\frac{1+y}{A(y)}d\varphi_1\right)^2\\
&+\frac{\mathcal{R}^2}{(x-y)^2}A(x)\left(-\frac{B(y)}{A(y)}d\varphi_1^2-\frac{dy^2}{B(y)}+\frac{dx^2}{B(x)}+\frac{B(x)}{A(x)}d\varphi_2^2\right)\ ,
\end{aligned}
\end{equation}
where 
\begin{equation}
    A(z)=1+\lambda z\ ,\qquad B(z)=(1-z^2)(1+\nu z)\ ,\qquad C=\sqrt{\lambda(\lambda-\nu)\frac{1+\lambda}{1-\lambda}}\ ,
\end{equation}
and $0<\nu\leq \lambda<1$ are dimensionless parameters, while $\mathcal{R}$ is a dimensionful radius. To avoid naked conical singularities in Eq.~\eqref{eq:BRoriginalMetric}, one needs to impose
\begin{equation}\label{eq:BRequilibrium}
    \lambda=\frac{2\nu}{1+\nu^2}\ ,
\end{equation}
and since $\lambda$ and $\nu$ are parameters related to the shape and angular momentum of the ring, the physical interpretation of such condition is to require that the centrifugal force exactly compensate the gravitational self-attraction. Therefore, one can refer to Eq.~\eqref{eq:BRequilibrium} as the equilibrium condition. As explained in the main text, we will not enforce such condition, so the black ring solution is described by three parameters.

Our goal now is to move from ring coordinates to harmonic coordinates. In order to do so, we need an intermediate step in which we simply consider asymptotically flat coordinates through the transformation~\cite{Heynen:2023sin}
\begin{equation}
\begin{gathered}
    x=-\left(\frac{1-\lambda}{1-\nu}\right)\frac{r^2-2\left(\frac{1-\lambda}{1-\nu}\right)\mathcal{R}^2\cos^2(\theta)}{r^2}\ , \qquad y=-\left(\frac{1-\lambda}{1-\nu}\right)\frac{r^2+2\left(\frac{1-\lambda}{1-\nu}\right)\mathcal{R}^2\sin^2(\theta)}{r^2}\ ,\\
    (\varphi_1, \varphi_2)=\frac{\sqrt{1-\lambda}}{1-\nu} (\phi_1, \phi_2)\ ,
    \end{gathered}
\end{equation}
in which one may now express the metric in terms of the coordinates $(t, r, \theta, \phi_1, \phi_2)$. Furthermore, it is convenient to make  the dependence on the mass $m$ and the angular momentum $J$ manifest in the black ring metric. Considering the asymptotic long-range behavior of the solution, the mass monopole and the spin dipole, \textit{i.e.} the mass and the spin itself, respectively read
\begin{equation}
    m=\frac{3\pi \mathcal{R}^2}{4G}\ ,\qquad J=\frac{\pi \mathcal{R}^3}{2G}\frac{\sqrt{\lambda(\lambda-\nu)(1+\lambda)}}{(1-\nu)^2}\ .
\end{equation}
Keeping $\lambda$ a free parameter, in the metric written in asymptotically flat coordinates we can now replace $(\mathcal{R}, \nu)$ in favor of $(m, J)$, having finally a metric described by three independent parameters $(m, J, \lambda)$. 

We can finally move to harmonic coordinates by considering the transformation~\eqref{eq:HarmonicCoordinatesD5}, that through the definition of harmonic Cartesian coordinates in Eq.~\eqref{eq:HarmCartesianD5}  have to satisfy the relation~\eqref{eq:BoxD5}.
The ansatz for the harmonic coordinate transformation is different from the Myers-Perry one since the topology is more complicated. Moreover, for the reasons explained in Sec.~\ref{sec:BlackRing}, we will consider only terms up to 1PM and second order in the spin, so that the coordinate transformation reads\footnote{Notice that at higher PM orders one should include logarithmic pieces as in the $D=5$ Myers-Perry case.}
\begin{equation}\label{eq:BRansatz}
\begin{aligned} 
    &r(R, \Theta)=R\sum_{i=0}^{1}\left(G m\rho(R)\right)^i\sum_{j=0}^{1}\sum_{k=0}^{2}\mathcal{C}_{i, 2j, 2k}^{(R)}(\lambda)\left(\frac{a}{R}\right)^{2j}P_{2k}(\cos\Theta)+O(G^2, a^3)\ , \\
    &\cos\theta(R, \Theta)=\cos\theta\sum_{i=0}^{1}\left(G m\rho(R)\right)^i\sum_{j=0}^{1}\sum_{k=0}^{2}\mathcal{C}_{i, 2j, 2k}^{(\Theta)}(\lambda)\left(\frac{a}{R}\right)^{2j}P_{2k}(\cos\Theta)+O(G^2, a^3)\ ,
\end{aligned}
\end{equation}
where now the coefficients that have to be fixed will be functions of the free parameter $\lambda$. Notice that among all the constraints to impose to this ansatz, we have to ensure that the monopole term in the space-part of the metric is independent of $\lambda$, since it must carry only the information of the physical mass. This extra constraint, which is not a priori satisfied from Eq.~\eqref{eq:BRansatz}, fixes to two the total number of gauge redundant parameters at 1PM, exactly as the previous cases. 

Finally, considering the explicit expression of Eq.~\eqref{eq:BRansatz} (reported in the Mathematica file \cite{AnchillaryFiles}), the black ring metric in harmonic coordinates (in a suitable gauge), in which we are neglecting quadrupoles that are not spin-induced, reads
\begin{equation}
    \begin{aligned}
    g_{tt}&=1-\frac{8 G m}{3\pi R^2}+\frac{12 G m a^2\lambda\cos (2 \Theta )}{\pi R^4(1+\lambda)}+O(G^2, a^3)\ ,\\
    g_{t\Phi_1}&=\frac{4 a G m \sin^2(\Theta)}{\pi R^2}+O(G^2, a^3)\ ,\\
    g_{RR}&=-1-\frac{4Gm}{3\pi R^2}+\frac{6Gma^2\lambda\cos(2\Theta)}{\pi R^4(1+\lambda)}+O(G^2, a^3)\ ,\\
    g_{\Theta\Theta}&=-\frac{4Gm}{3\pi}-R^2+\frac{2Gma^2(1+3\lambda\cos(2\Theta))}{\pi R^2(1+\lambda)}+O(G^2, a^3)\ ,\\
    g_{\Phi_1\Phi_1}&=-\frac{4Gm\sin^2(\Theta)}{3\pi}-R^2\sin^2(\Theta)+\frac{Gma^2\sin^2(\Theta)}{\pi R^2}\Big(-1+3(1+3\lambda)\cos(2\Theta)\Big)+O(G^2, a^3)\ ,\\
    g_{\Phi_2\Phi_2}&=-\frac{4Gm\cos^2(\Theta)}{3\pi}-R^2\cos^2(\Theta)+\frac{Gma^2\cos^2(\Theta)}{\pi R^2}\Big(-1+3(-1+3\lambda)\cos(2\Theta)\Big)+O(G^2, a^3)\ .\\
    \end{aligned}
\end{equation}
This metric can be now compared directly with the amplitude-based one, obtaining a perfect match up to the order we are considering.

\bibliography{biblio}

\end{document}